\documentclass{article}

\usepackage{tabularx}   
\usepackage{microtype}
\usepackage{graphicx}
\usepackage{subfigure}
\usepackage{booktabs} 
\usepackage{adjustbox}
\usepackage{ulem}
\usepackage{multirow}
\usepackage{array}
\usepackage{amssymb}
\usepackage{longtable}
\usepackage{enumitem}
\usepackage[T1]{fontenc}
\usepackage{hyperref}

\usepackage[preprint]{icml2026}
\usepackage{amsmath}
\usepackage{amssymb}
\usepackage{mathtools}
\usepackage{amsthm}
\usepackage{makecell}

\usepackage[capitalize,noabbrev]{cleveref}

\theoremstyle{plain}

\theoremstyle{definition}

\theoremstyle{remark}

\usepackage[textsize=tiny]{todonotes}

\theoremstyle{plain}
\theoremstyle{definition}
\theoremstyle{remark}

\usepackage[textsize=tiny]{todonotes}

\icmltitlerunning{The Spectral Amplitude Principle for Dynamics of Quantum Neural Networks}

\begin{document}

\twocolumn[
    \icmltitle{The Spectral Amplitude Principle for Dynamics of Quantum Neural Networks}
    
    
    
    \icmlsetsymbol{equal}{*}
    \begin{icmlauthorlist}
    \icmlauthor{Yi-hang Xu}{xxx}
    \icmlauthor{Dan-Bo Zhang}{yyy}
    \icmlauthor{Junchi Yan}{xxx}
    \end{icmlauthorlist}
    
    \icmlaffiliation{xxx}{School of Artificial Intelligence and School of Computer Science, Shanghai Jiao Tong University, Shanghai, China}
    \icmlaffiliation{yyy}{School of Physics, South China Normal University, Guangzhou, China} 
    
    \icmlcorrespondingauthor{Dan-Bo Zhang}{dbzhang@m.scnu.edu.cn}
    \icmlcorrespondingauthor{Junchi Yan}{yanjunchi@sjtu.edu.cn}
    
    \icmlkeywords{Machine Learning, ICML}
    
    \vskip 0.3in
]

\printAffiliationsAndNotice{}



\begin{abstract}
 The mechanism governing the training dynamics of Quantum Neural Networks (QNNs) remains under-explored. In classical Deep Neural Networks (DNNs), training is dominated by ``Spectral Bias,'' i.e. prioritizing learning low-frequency components and struggle for high-frequency details. In this work, we theoretically and empirically identify a distinct mechanism in QNNs, which we term \textit{Spectral Amplitude Priority}. By analyzing the frequency-domain gradients and residual dynamics via the Quantum Neural Tangent Kernel (QNTK), we prove that QNN training is governed primarily by the magnitude of spectral components rather than their frequency indices. Consequently, QNNs can efficiently capture high-frequency functions—provided they have significant amplitude—thereby overcoming the inherent limitations of their classical counterparts. We validate this principle on both synthetic high-frequency functions and quantum-advantage tasks. The results show that QNNs significantly outperform DNNs in high-frequency tasks, offering an explanation for QNNs' superior expressivity in complex spectral landscapes.

\end{abstract}
\section{Introduction}
\label{sec:level1}
Quantum machine learning (QML), often in the form of quantum neural networks (QNNs)~\cite{Dunjko_2018} and embodied as parameterized quantum circuits (PQC)~\cite{Benedetti_2019} has received intensive studies in recent years. In particular, efforts have been made to explore the quantum advantages for QML,  in different aspects e.g. complexity~\cite{2020Power,PhysRevLett.124.010506,2020A},  expressibility~\cite{PhysRevLett.128.080506,PhysRevA.103.032430}, generalization~\cite{caro2022generalization} and trainability~\cite{PRXQuantum.2.040337,cerezo2021cost}. 



One bottleneck of adopting QNNs is its notorious difficulty in training, where the so-called barren plateaus~\cite{mcclean2018barren} problem remains open, despite there have been various mitigating methods devised: structured ansatz design~\cite{zhang2020toward,uvarov2021barren}, initialization strategies~\cite{zhang2022escaping,park2024hamiltonian}, and derivative-free optimization~\cite{xie2025quantum,nadori2025batched,acosta2024adiabatic}.  

In this paper, we try to address another fundamental yet unresolved problem, i.e. understanding the training dynamics of QNNs. It not only relates to the aforementioned training stability problem, but also sheds light on the expressibility of QNNs. For instance, whether or in what conditions, the QNN can capture the high-frequency spectral components of training data which is otherwise very difficult for its classical counterpart i.e. deep neural networks (DNNs) as already widely discussed in literature. Specifically, DNNs are known inherently limited by their low-frequency spectral preference~\cite{xu2018frequency,rahaman2019spectral}. In this work, we provide a systematic answer through both theoretical analysis and empirical evidence, showing that QNNs can enjoy better capability in capturing high-frequency components\footnote{For conciseness of analysis, the QNNs studied in this paper are grounded in the data re-uploading architecture~\cite{P_rez_Salinas_2020} i.e. composed of an interleaved data encoding block and a trainable circuit block which is widely used in QNN literature.}. These results, to our best knowledge, have not been reported in literature, although there are some preliminary analyses~\cite{jaderberg2024let,poppel2025mitigating} from the frequency perspective.

\begin{table*}[t!]
\centering
\caption{Related works on spectral analysis in DNNs and QNNs.}
\begin{adjustbox}{max width=\textwidth}
\begin{tabular}{p{0.8cm} p{3.2cm} p{4.4cm} p{2.5cm} p{5.4cm}}
\toprule
Domain & Reference & Framework & Spectral Behavior & Primary Focus \\
\midrule
\multirow{15}{*}{DNNs} 
 & ~\cite{arpit2017closer} & Empirical & \multirow{15}{*}{Spectral Bias} & Memorization \& generalization phases \\
 & ~\cite{rahaman2019spectral} & Fourier Analysis & & Robustness to perturbations and convergence speed \\
 &\cite{zhang2019explicitizing} & Linear Frequency Principle dynamics  & & Predicting learning outcomes via dynamics models \\
 &\cite{luo2022exact} & Linear Frequency Principle dynamics & & Linking error decay rate to activation functions \\
 & \cite{basri2020frequency} & Neural Tangent Kernel & & Dynamics under non-uniform data distributions \\
 & \cite{ijcai2021p304} & Neural Tangent Kernel & & Mathematically linking between eigenvalues and convergence \\
 & \cite{molina2024understanding} & Partial Differential Equations & & Role of initialization distribution in frequency bias \\
\midrule
\multirow{9.5}{*}{QNNs} 
 & \cite{jaderberg2024let} & Trainable Frequency Modeling & Spectral Gap Modulation & Aligning with target spectrum \\
 & \cite{heimann2025learning} & Fourier Series Approximation & Truncated Spectrum & Evaluating ansatz capability for Fourier series \\
 & \cite{poppel2025mitigating} & Frequency Select Algorithms & Non-unique Frequency Map & Mitigating trainability issues via frequency selection \\
 \cmidrule(l){2-5}
 & \textbf{Our Work} & {Analysis of gradient \& residuals dynamics with QTNK} & {Spectral Amplitude Priority} & {fitting priority is by spectral magnitude} \\
\bottomrule
\label{compare}
\end{tabular}
\end{adjustbox}
\label{tab:compare}
\end{table*}

Specifically, we show that QNNs follow a distinctive learning trajectory characterized by spectral amplitude priority. Our analysis is facilitated by establishing a gradient analysis framework to characterize the training mechanism. To further rigorously quantify the dynamics, we incorporate the quantum neural tangent kernels~(QNTK)~\cite{PhysRevLett.130.150601,PRXQuantum.3.030323} to derive an evolution equation for the training residual dynamics in the frequency domain. 

Theoretically, we prove that (in Sec.~\ref{sec:QNTK}) when the learning rate is reasonably small e.g. 0.01, the residuals corresponding to the high-amplitude spectral components decay quickly (i.e. exponentially) over training epochs\footnote{Empirically, although slight discrepancies are observed between the analytical solution and actual evolution, for low-amplitude components during the initial training phase, the proposed residual dynamics framework yields a robust approximation of the overall training trajectory.}.  In contrast, it is shown that for classical DNNs, the model is always confined to fit low-frequency information first, and it is very challenging to capture the high-frequency part~\cite{xu2018frequency,rahaman2019spectral}. These results indicate an advantage of QML as QNNs can enjoy superior expressiveness to fit with arbitrary frequency spectrum of the training data, as long as its corresponding amplitude is strong enough. \textbf{Our findings can be summarized as follows:}

\noindent \textbullet~ \textbf{Spectral Amplitude Priority Principle for QNNs:} We uncover the distinction between DNNs' known spectral bias (also called the frequency principle~\cite{xu2018frequency} i.e. favoring fitting the low-frequency part first in training) and QNNs' spectral amplitude principle namely the fitting is governed by the signal amplitude rather than frequency indices.

\noindent \textbullet~ \textbf{Expressiveness Advantages of QNNs:} High-frequency signals can be effectively learned when the corresponding spectral amplitude is strong, which indicates the potential advantage of QNNs. We further derive the closed-form expression for the gradients of the training loss in the frequency domain, providing explicit conditions for the frequency components prioritized during (early) training.
    
\noindent \textbullet~ \textbf{Residual Dynamics in Frequency Domain:} We reformulate the QNTK in frequency domain to derive a differential equation governing the training dynamics of residuals. This analytical framework provides a rigorous  understanding that the residuals decay exponentially over training time.
    
\noindent \textbullet~ \textbf{Verification on Quantum Advantage Tasks:} Experiments on tasks with theoretically provable quantum advantages w.r.t. complexity and expressibility are performed. The results validate the existence of the spectral amplitude priority, as well as the performance superiority over their classical counterparts on simulation environments. 

\textbf{Remarks and Limitations.} Existing studies~\cite{jaderberg2024let,poppel2025mitigating} for the training of QNNs are mostly confined to ideal quantum circuits as the noise can be complex which renders the theoretical analysis elusive. Likewise, we leave the experiments and further theoretical analysis for specific hardware environments for future work. As will be shown in detail in Sec.~\ref{sec:method}, our framework rests on two key assumptions. First, Our empirical validation relies on classical simulations of ideal state evolution. On real quantum processors, hardware constraints such as limited connectivity, gate infidelity, and compilation overhead may introduce deviations. Second, the Amplitude Priority mechanism is strictly derived from a gradient-based optimization perspective. Conversely, gradient-free optimization strategies, e.g., SPSA or Nelder-Mead may call for separate investigations. Table~\ref{tab:compare} categorizes existing literature and see Appendix~\ref{sec:level2} for details of related work.

\section{Methods}
\label{sec:method}
We elucidate the Spectral Amplitude Principle in QNNs. First, we establish the gradient analysis framework to characterize the training mechanism in Sec.~\ref{sec:spec_amp}. Subsequently, we propose the QNTK in the frequency domain, and then derive the governing evolution equation for the residual dynamics in Sec.~\ref{sec:dynamics_freq}, providing a rigorous theoretical guarantee for the observed convergence behavior.
Table~\ref{tab:symbols} in Appendix provides a list of notations used to describe the QNN, the data re-uploading mechanism, frequency-domain analysis, and the residual analysis.

\subsection{The Spectral Amplitude Principle for QNNs}
\label{sec:spec_amp}
\subsubsection{Gradient Analysis in Frequency Domain}\label{sec:gradient}
We focus on univariate function approximation to investigate the training dynamics of QNNs. Note that in this paper, our analysis is primarily conducted within the framework of the ideal circuit model, abstracting away hardware-specific overheads such as error correction and gate noise. 

\begin{figure}[tb!]
\centering
 \includegraphics[width=0.42\textwidth]{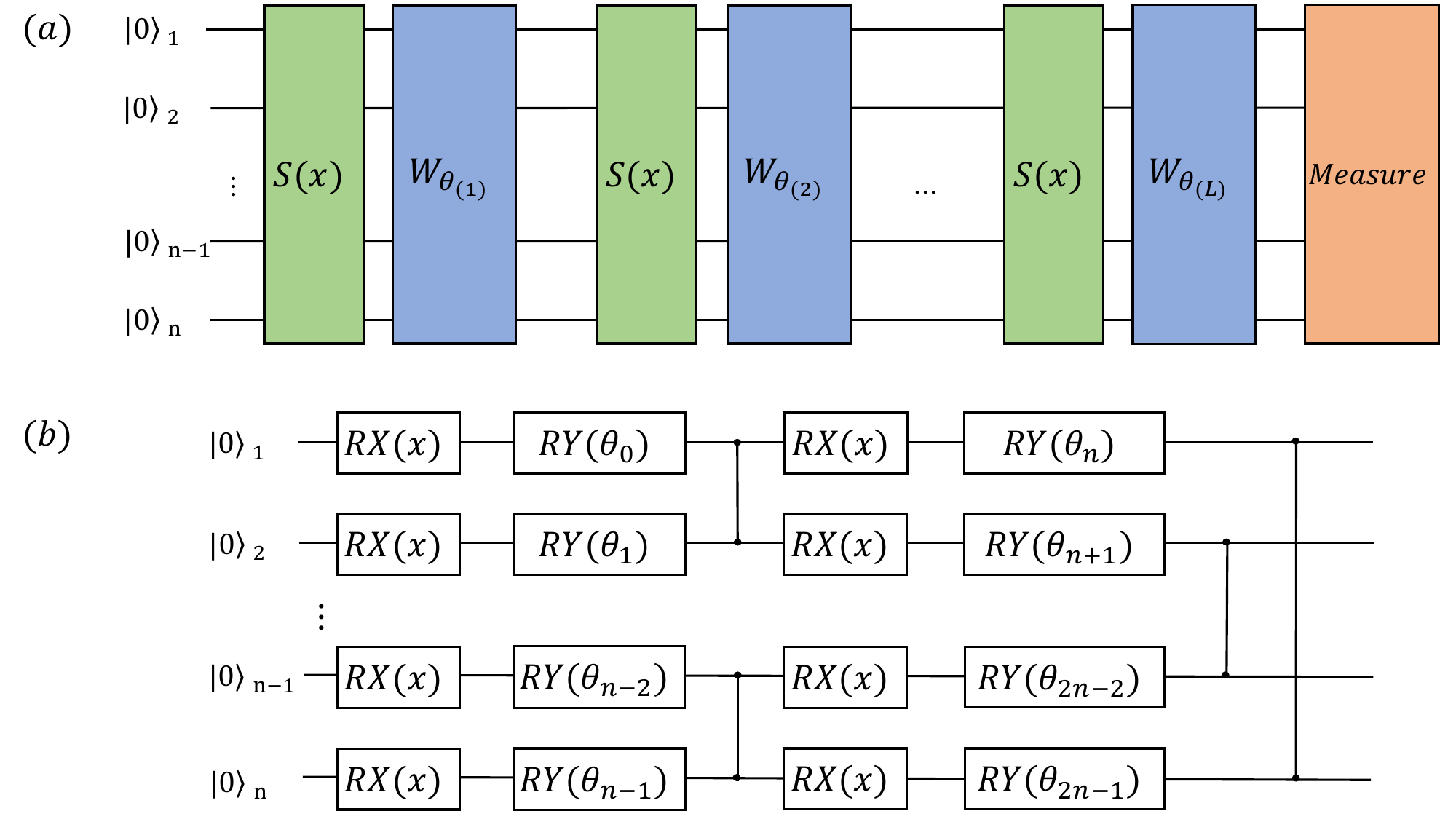}
 \caption{\textbf{The scheme of QNNs studied in the paper.} (a)  $S(x)$ is the data encoding circuit block and $W_{\theta _{(p)}}$ is the trainable circuit block. (b) One of the layers of $S(x)$ and $W_{\theta_{(p)}}$ specific gates and their dependence on $x$, $\theta$. The encoding circuit $S(x)$ embeds classical data into the rotation angles of RX gates, thereby encoding classical information into quantum state. }\label{Fig:qc}
 \vskip 0.1in
\end{figure}

Without loss of generality, we follow the literature in QNNs training analysis,  by employing the data-reuploading scheme~\cite{P_rez_Salinas_2020} that alternatively sets the data encoding circuit block $S(x)$ and the trainable circuit block $W(\boldsymbol{\theta} )$. As shown in Fig.~\ref{Fig:qc}, the data encoding block consists of gates of the form $S(x) = e^{-ix\mathit{H}}$, where $\mathit{H}$ is an arbitrary Hamiltonian. $\boldsymbol{\theta}  = (\theta_{(1)} ,\theta_{(2)},\dots ,\theta_{(L)})$ is the set of trainable parameters. The $L$-layer QNN is
\begin{eqnarray}
U(x,\boldsymbol{\theta} ) =  W_{\theta_{(L)}}S(x) \cdots W_{\theta_{(2)}}S(x) W_{\theta_{(1)}} S(x).
\label{eq:U}
\end{eqnarray}

The output state provides a way to express a function on $x$, where $M$ is the measurement operator.
\begin{eqnarray}
f(x,\boldsymbol{\theta}) = {^{n\otimes}}{\left \langle  0| U^{\dagger } (x,\boldsymbol{\theta} )MU(x,\boldsymbol{\theta} )|0\right \rangle}{^{\otimes n}}.
\label{eq:eigen}
\end{eqnarray}
The parameter $\theta$ is determined by minimizing the loss  $L(\boldsymbol{\theta})$.

To formally analyze the training dynamics in frequency domain, the QNN output is expressed as a partial Fourier series~\cite{PhysRevA.103.032430,Casas_2023}:
\begin{eqnarray}
\hat{f}(k,\boldsymbol{\theta}) = \sum_{\omega \in \Omega }^{} {C(\omega ,{\boldsymbol{\theta}})e^{i\mathbf{\omega x}}},
\label{eq:flyjsk}
\end{eqnarray}
where the data encoding circuit determines the frequency $\omega$ and the rest of the circuit determines the coefficients $C(\omega,\boldsymbol{\theta})$ of the summation, and $\Omega$ denotes any accessible frequency as determined by the data encoding circuit module $S(x)$. 

By Fourier transformation, we have
\begin{eqnarray}
\hat{f}(k,{\boldsymbol{\theta}}) = \sum_{\omega \in \Omega }^{} {C(\omega ,{\boldsymbol{\theta}}) \delta_{k,\omega}},
\label{eq:flyjsk}
\end{eqnarray}
where the Dirac $\delta$ indicates that $\hat{f}(k,{\boldsymbol{\theta}})$  have a non-zero value only when $k$ belongs to the accessible frequency. 

As a generalization to \cite{rahaman2019spectral,Xu_2020} in the classical setting, $\hat{y}(k)$ denotes the frequency components of the target function to learn, and the frequency-domain residual is defined as the difference between the QNN output spectrum and the target spectrum:
\begin{equation}
\hat{\varepsilon}(k,{\boldsymbol{\theta}}) = \hat{f} (k,\theta)-\hat{y}(k),     
\label{eq:residule}
\end{equation}
We get the gradient of loss:
\begin{eqnarray}
\frac{\partial \hat{L}(k)}{\partial { \theta _{j}} }  =   \hat{\varepsilon}(k,{\boldsymbol{\theta}}) \frac{\partial \hat{f}^{\dagger}(k,{\boldsymbol{\theta}})}{\partial { \theta _{j}} } + h.c.
\label{eq:tidu}
\end{eqnarray}
where h.c. denotes the Hermitian conjugate of the preceding terms. The residual can be written as $\hat{\varepsilon}(k,\boldsymbol{\theta}) = A(k)e^{i\phi (k)}$:
\begin{eqnarray}\label{eq:gradient_k}
\frac{\partial \hat{L}(k)}{\partial { \theta _{j}} } = A(k)\left[e^{i\phi (k)}\sum_{\omega \in \Omega }^{} \frac{\partial C^{\dagger}(\omega ,{\boldsymbol{\theta}}) }{\partial { \theta _{j}} } \delta_{k,\omega}+h.c.\right].
\end{eqnarray}


The analytical expression in Eq,~\ref{eq:gradient_k} provides a fundamental insight into the training dynamics of QNNs. Unlike classical DNNs, where the gradient magnitude typically decays with frequency (e.g., scaling as $\frac{1}{k^2} $ due to the smoothness of activation functions), the gradient contribution in QNNs does not inherently penalize high-frequency components. 

Specifically, the derivative term $\frac{\partial \hat{f}}{\partial \theta}$ depends on the circuit structure and data encoding but remains \textit{spectrally neutral}—it does not vanish as the frequency index $k$ increases, provided $k$ is within the accessible spectrum $\Omega$ determined by the encoding strategy. Consequently, the magnitude of the gradient $\frac{\partial \hat{L}(k)}{\partial \theta_j}$ is linearly proportional to the residual amplitude $A(k)$. This implies that the optimizer will prioritize reducing the error of the frequency components with the largest current residuals ($A(k)$), regardless of whether $k$ represents a low or high frequency. We term this phenomenon \textbf{Spectral Amplitude Priority}, which fundamentally differentiates QNN optimization dynamics from the Spectral Bias observed in classical networks.

\subsubsection{Validation of Gradient Behavior}
\begin{figure}[tb!]
\centering
 \includegraphics[width=0.4\textwidth]{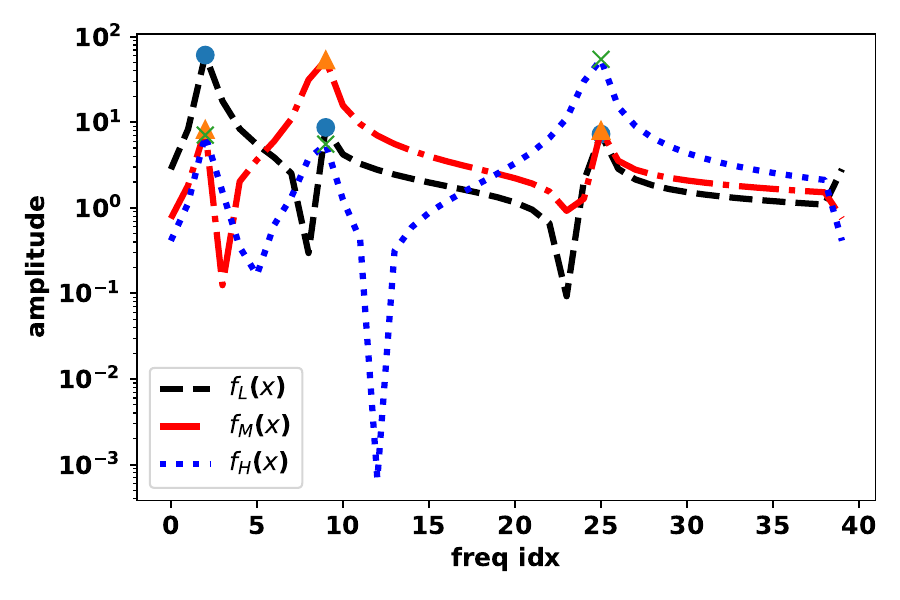}
\caption{ \textbf{The amplitude-frequency plots of the $f_L(x)$, $f_M(x)$, and $f_H(x)$} in Eq.~\ref{eq:curves}. The x-axis represents the index of each frequency component after applying the Fourier transform to the function, while the y-axis denotes the amplitude. The black, red, and blue curves correspond to the frequency and amplitude distributions of $f_L(x)$, $f_M(x)$, and $f_H(x)$, respectively.} \label{Fig:amp}
\vskip 0.1in
\end{figure}

We follow the experimental setup as described in Sec.~\ref{sec:exp}. To investigate the response to different spectral amplitude compositions, we select three target functions dominated by low, middle, and high frequencies, respectively:
\begin{equation}
\begin{split}
&f_L(x) = 0.9\sin(x) +0.1\sin(3x)+0.1\sin(8x),\\
&f_M(x) = 0.1\sin(x) +0.9\sin(3x)+0.1\sin(8x),  \\
&f_H(x) = 0.1\sin(x) +0.1\sin(3x)+0.9\sin(8x). 
\label{eq:curves}
\end{split}
\end{equation}
The amplitude-frequency plots of the three functions are given in Fig.~\ref{Fig:amp}. Since frequencies of the training data are discrete and non-uniformly distributed, we have replaced the physical frequencies with their corresponding indices, denoted as ``freq idx" (larger index  higher frequency). 

The quantum circuit is configured with 4 qubits and 20 layers, comprising a total of 160 trainable parameters.

\begin{figure*}[tb!]
 \begin{minipage}{0.32\linewidth}    \centerline{\includegraphics[width=\textwidth]{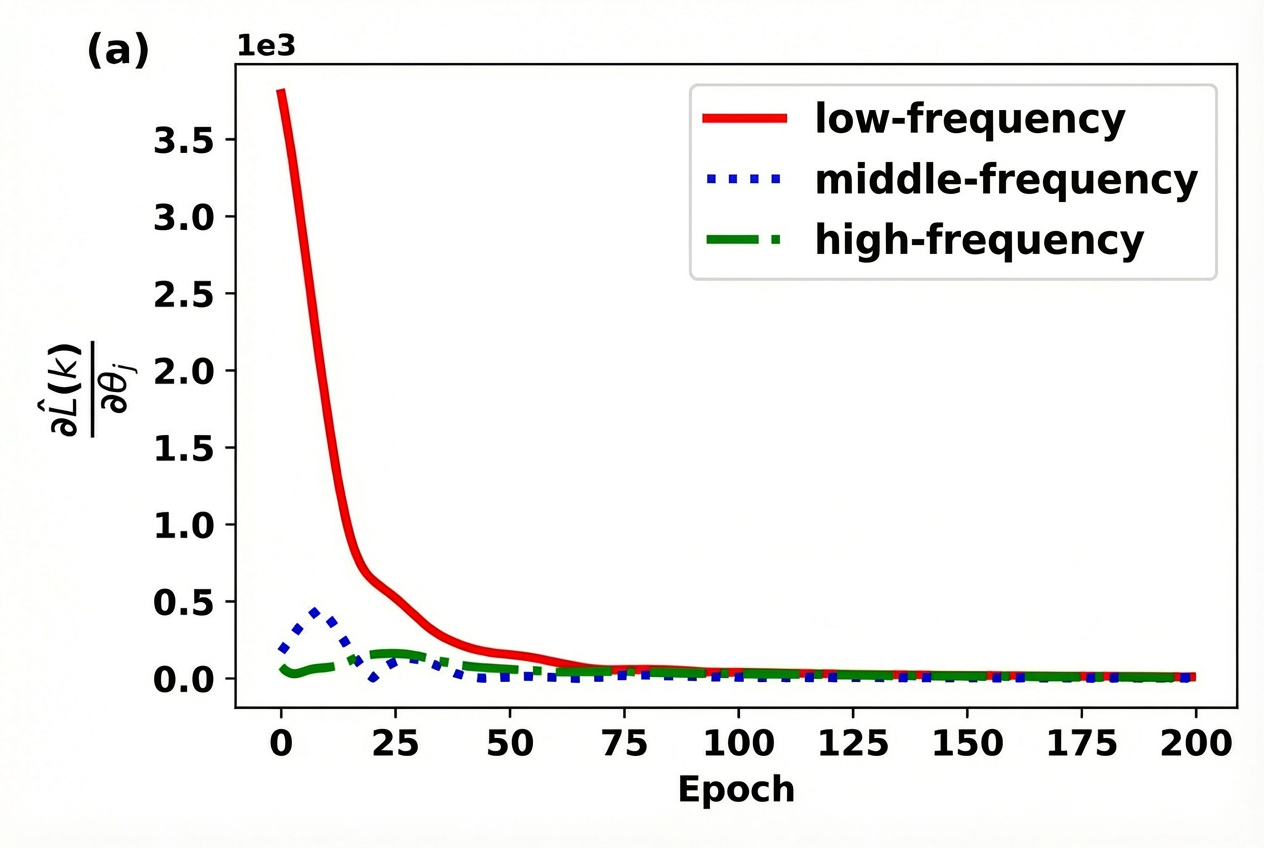}}
 \end{minipage}
 \begin{minipage}{0.32\linewidth}    \centerline{\includegraphics[width=\textwidth]{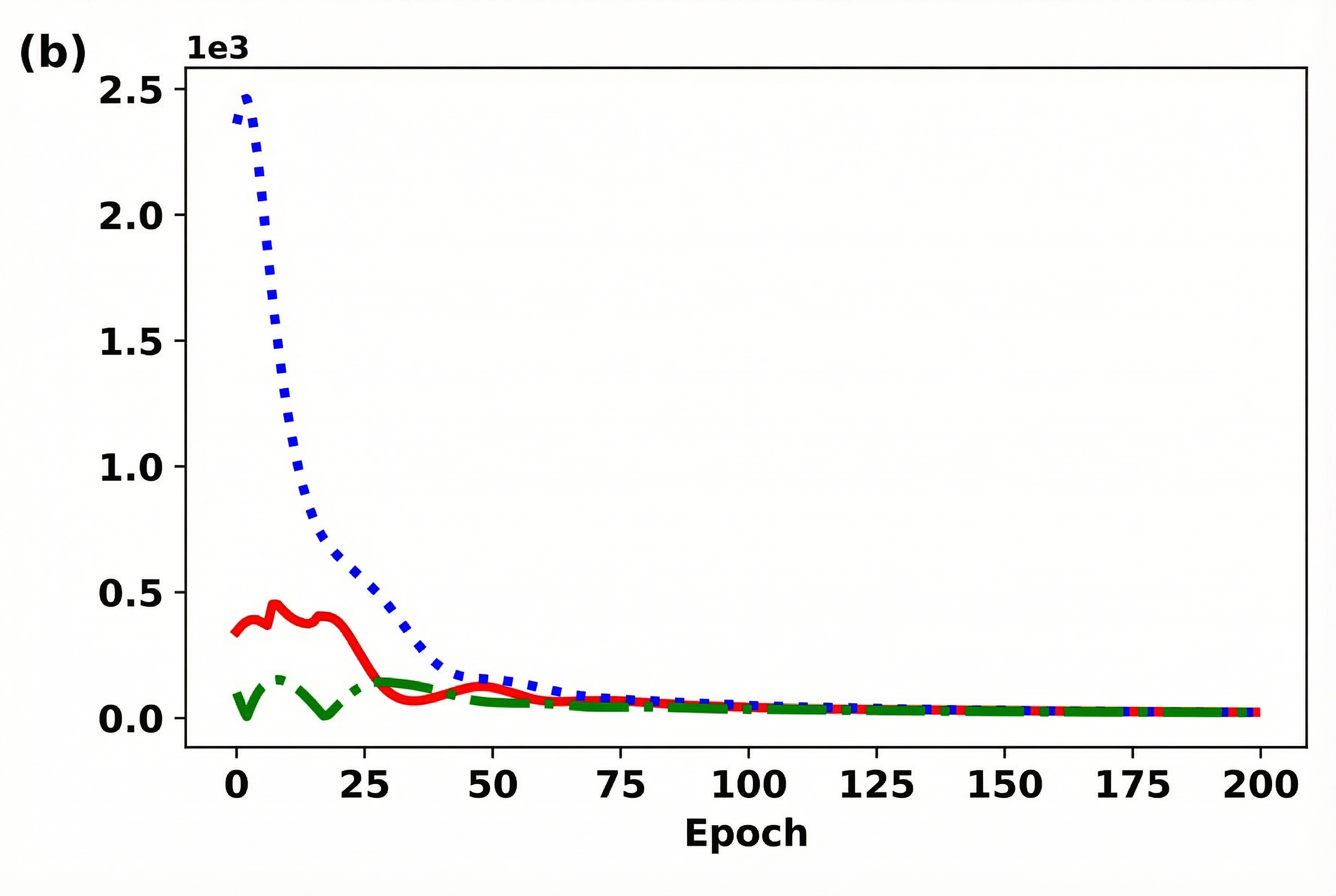}}
 \end{minipage}
  \begin{minipage}{0.32\linewidth}
    \centerline{\includegraphics[width=\textwidth]{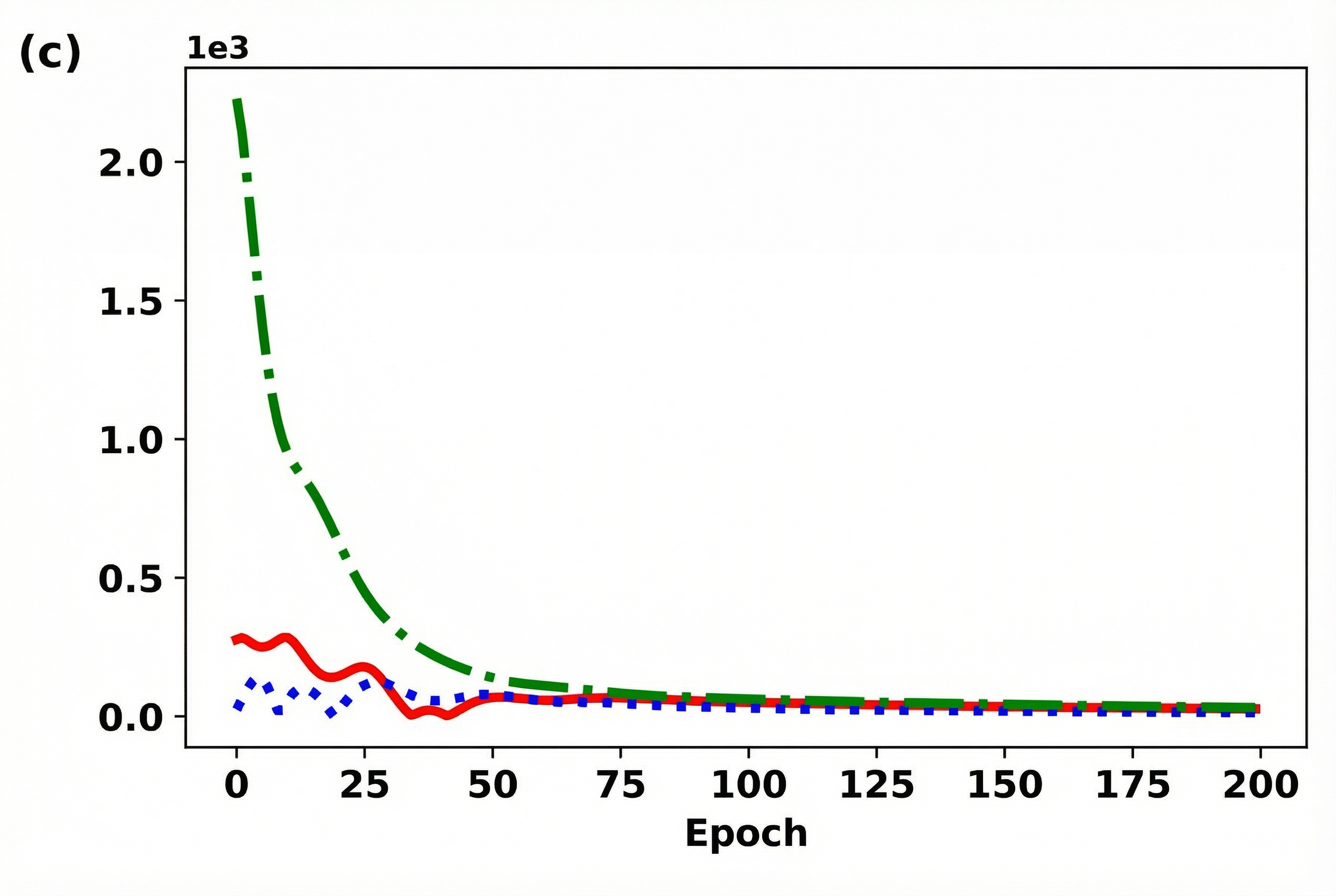}}
 \end{minipage}
    	\caption{Evolutions of gradients derived from Eq.~\ref{eq:gradient_k} for loss functions when training QNNs for fitting one-variable curves of low-frequency~(a), middle-frequency~(b) and high-frequency~(c) dominated functions. }\label{Fig: L_K}
\vskip 0.1in
\end{figure*}

We systematically investigate the convergence behavior of the loss gradients corresponding to distinct frequency components during training for these three curves. It is crucial to note that the gradient dynamics visualized in Fig.~\ref{Fig: L_K} are explicitly computed and plotted according to the analytical form derived in Eq.~\ref{eq:gradient_k}, rather than merely raw numerical outputs from an optimizer.

The results show that QNNs adaptively prioritize the convergence of frequency components with high amplitudes. Specifically, when the target function is high-frequency dominated ($f_H(x)$), the gradients for high-frequency components—computed via Eq.~\ref{eq:gradient_k}—exhibit significant magnitudes immediately upon initialization, thereby circumventing the spectral suppression typically observed in DNNs. This confirms that the training dynamics of QNNs are fundamentally governed by amplitude priority, providing robust empirical support for the Spectral Amplitude Principle.

\begin{figure*}[tb!]
 \begin{minipage}{0.32\linewidth}
 	\centerline{\includegraphics[width=\textwidth]{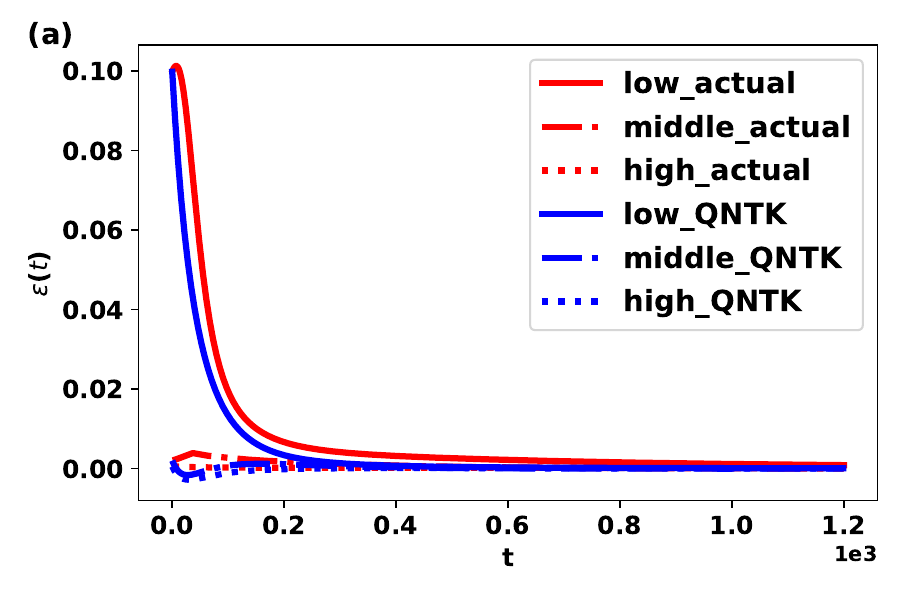}}
 \end{minipage}
 \begin{minipage}{0.32\linewidth}
     \centerline{\includegraphics[width=\textwidth]{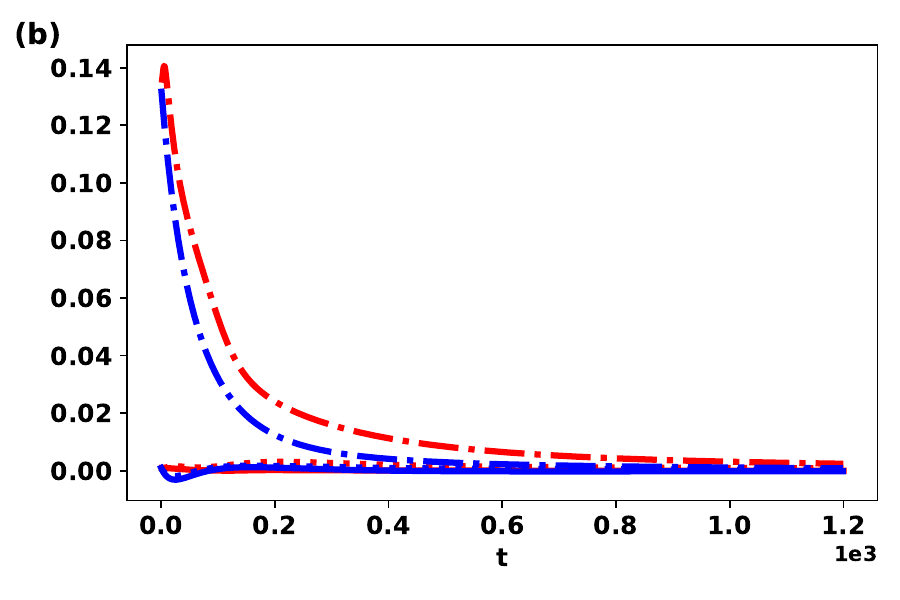}}
 \end{minipage}
  \begin{minipage}{0.32\linewidth}
     \centerline{\includegraphics[width=\textwidth]{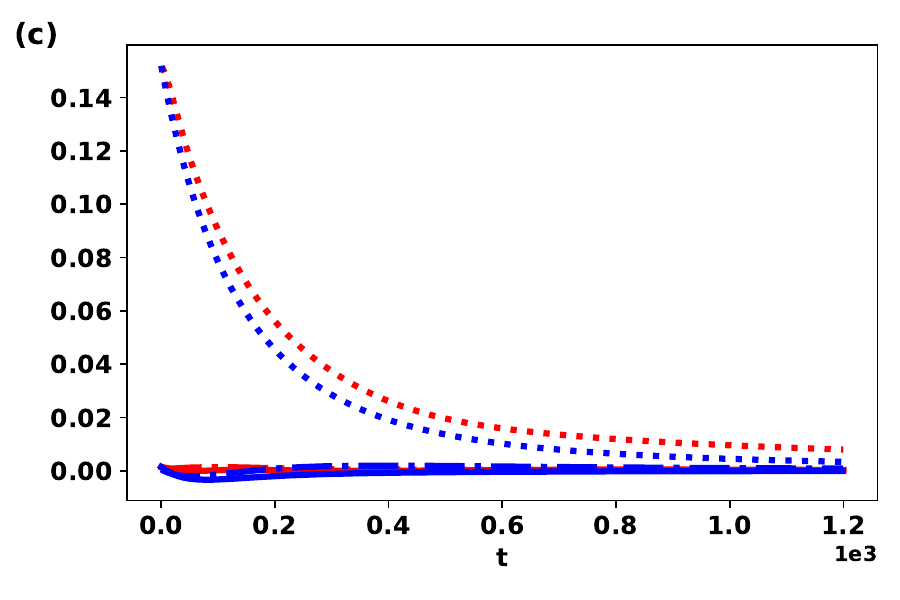}}
 \end{minipage}
	\caption{\textbf{Comparison of residual dynamics between actual ones and predicted by QNTK when learning low-frequency~(a), middle-frequency~(b) and high-frequency~(c) dominated functions.} We plot the actual dynamics of $\varepsilon(t)$ for the above three functions, taking three peak frequencies in frequency domain corresponding to low, medium, and high frequencies, and their $\varepsilon(t)$ theoretical predictions of QNTK. The qubit number is $8$ with learning rate $\eta=0.01$. Parameters in the ansatz are initialized randomly in [$0$, $2\pi$].}\label{Fig:fit}
\vskip 0.1in
\end{figure*}
\subsection{Residual Dynamics in Frequency Domain}
\label{sec:dynamics_freq}
We now investigate the training dynamics by capturing the average behavior of the residual dynamics at different frequencies in the frequency domain.  
\subsubsection{Residual Analysis Via QNTK}
\label{sec:QNTK}


In line with the setting~\cite{PhysRevLett.130.150601,PRXQuantum.3.030323} in $x$-space, the difference equation for the residual dynamics (Eq.~\ref{eq:residule}) derived from the loss gradient is expressed as follows, where the variable $\boldsymbol{\theta}$ is omitted for brevity: 
\begin{eqnarray}\label{eq:weifen}
\Delta \varepsilon(x_{i^{'} } ) { \approx \sum_{j}^{} \frac{\partial \varepsilon(x_{i^{'} } )}{\partial \theta_{j}} d\theta_{j} =  } -\sum_{i}^{}\eta \mathcal{K}_{\boldsymbol{\theta}}(x_{i^{'} },x_{i})\varepsilon  (x_{i } ),\nonumber 
\end{eqnarray}
where $\mathcal{K}_{\boldsymbol{\theta}}(x_{i^{'} }x_{i}) = \sum_{j}^{} \frac{\partial{d} \varepsilon  (x_{i^{'} } )}{\partial \theta _{j} } \frac{\partial{d} \varepsilon (x_{i } )}{\partial \theta _{j}} $. The derivation of the residual dynamics operates strictly within the lazy training regime~\cite{PhysRevLett.130.150601}. This theoretical framework relies on the assumption of a sufficiently small learning rate, which effectively confines the parameters to a small neighborhood of their initialization. Such an approximation allows the QNTK to be treated as a time-invariant constant during training, i.e., $ \mathcal{K}(x_{i^{'} },x_{i}) \approx \overline{\mathcal{K}} (x_{i^{'} },x_{i})$. Here 
\begin{eqnarray}
\overline{\mathcal{K}} (x_{i^{'} },x_{i})=
\frac{2L(D|\chi_{x_{i^{'}},x_{i}} |^{2}-1)}{(D^{2}-1)^{2}}(D\text{Tr}(M^2)-(\text{Tr}M)^2),\nonumber 
\end{eqnarray}

We can make Fourier transformation and write the difference equation for residual dynamics in the frequency domain as, 
\begin{eqnarray}\label{eq:weifenf}
\Delta\varepsilon (k ) = \sum_{i}^{} -\eta \overline{\mathcal{K}}(k,k')\varepsilon (k' ),
\end{eqnarray}
where the Fourier transform of QNTK is, 
\[\overline{\mathcal{K}}(k',k) = \frac{1}{N} \sum_{ii'}\overline{\mathcal{K}}(x_{i^{'} },x_{i})e^{ik'x_{i^{'}}  }e^{-ikx_{i}}. \]
For the above linear difference Eq.~\eqref{eq:weifenf}, we can directly solve the linear equation and finally get the solution as:
\begin{eqnarray}
\varepsilon (t) = e^{-\eta\overline{\mathcal{K}}t  } \varepsilon (0), 
\label{eq:fianl}
\end{eqnarray}
where $\varepsilon (0)$ denotes the residual at $t$ = 0.
\begin{figure*}[tb!]
 \begin{minipage}{0.32\linewidth}
 	\centerline{\includegraphics[width=\textwidth]{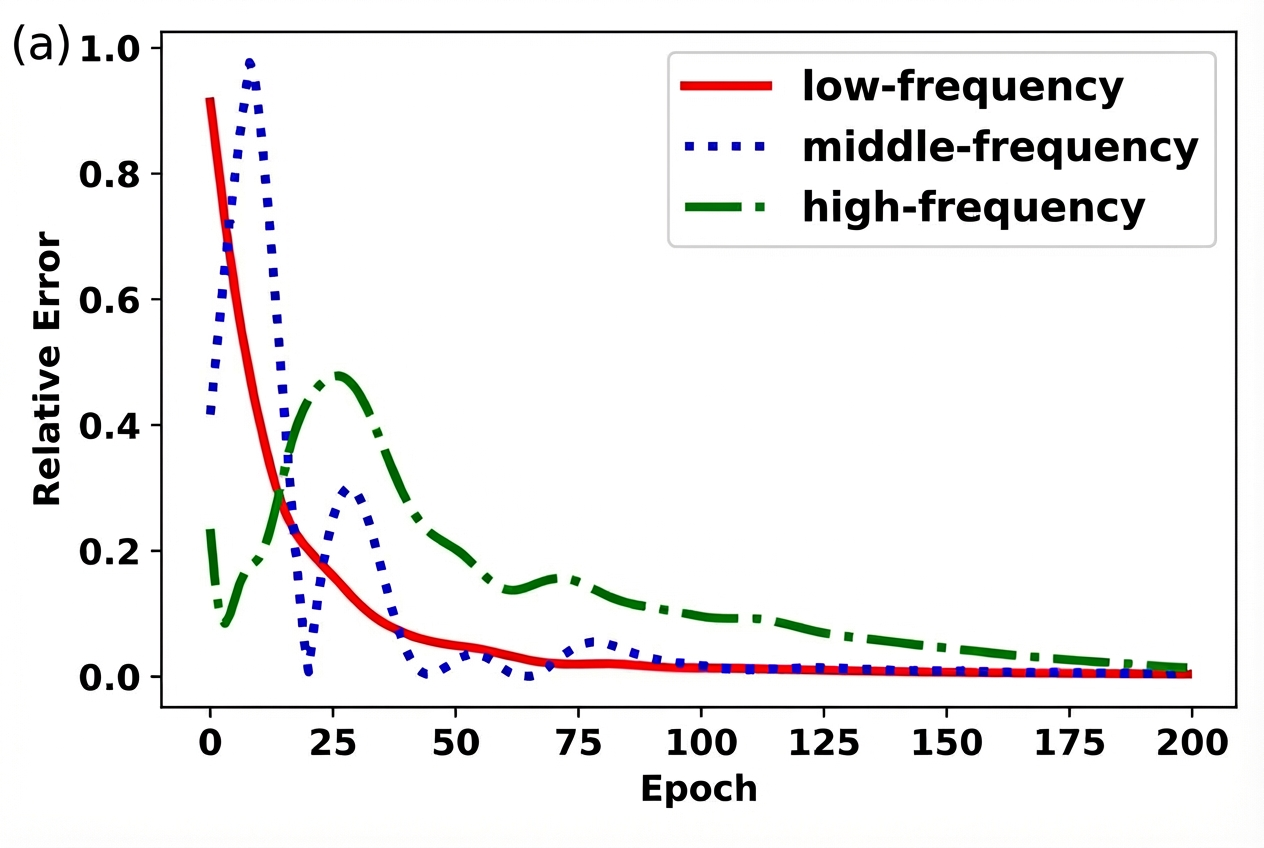}}
 \end{minipage}
 \begin{minipage}{0.32\linewidth}
     \centerline{\includegraphics[width=\textwidth]{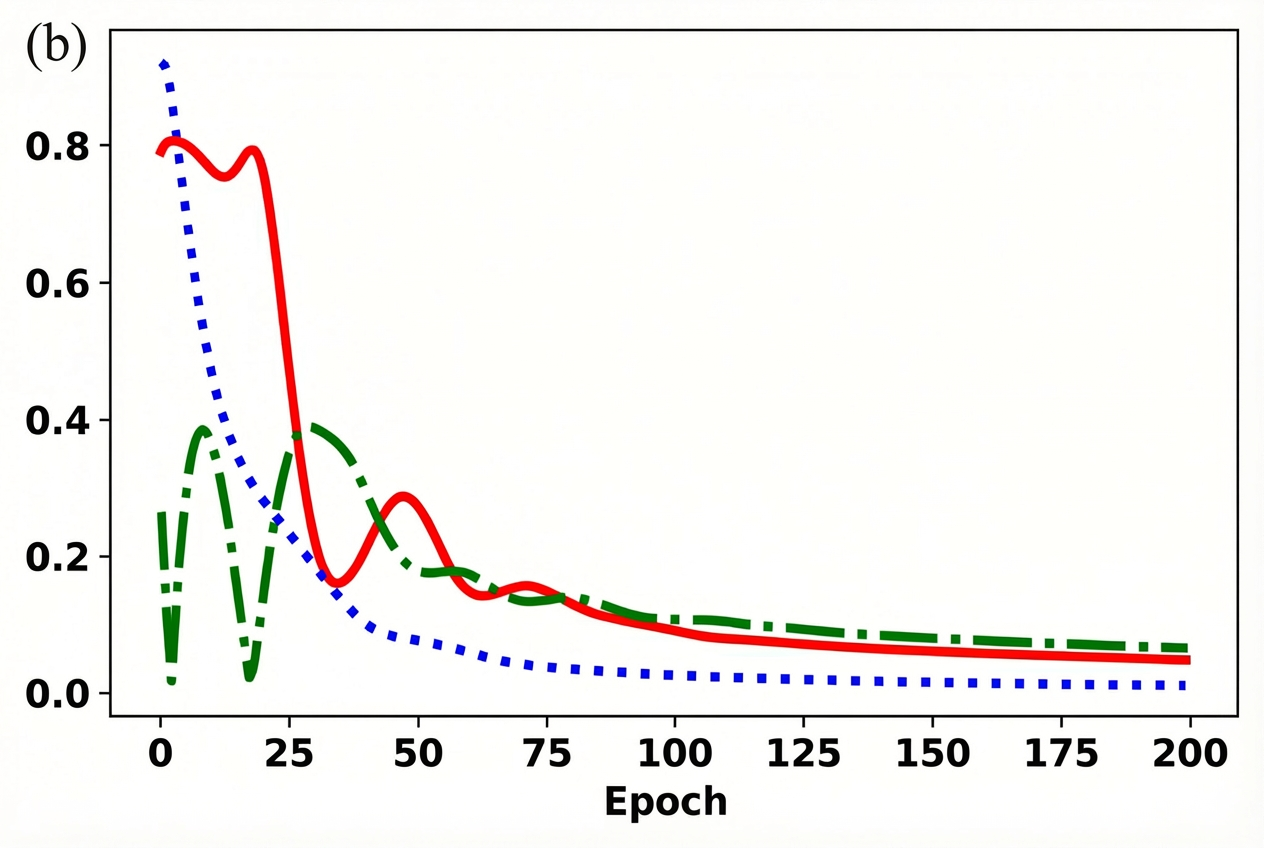}}
 \end{minipage}
  \begin{minipage}{0.32\linewidth}
     \centerline{\includegraphics[width=\textwidth]{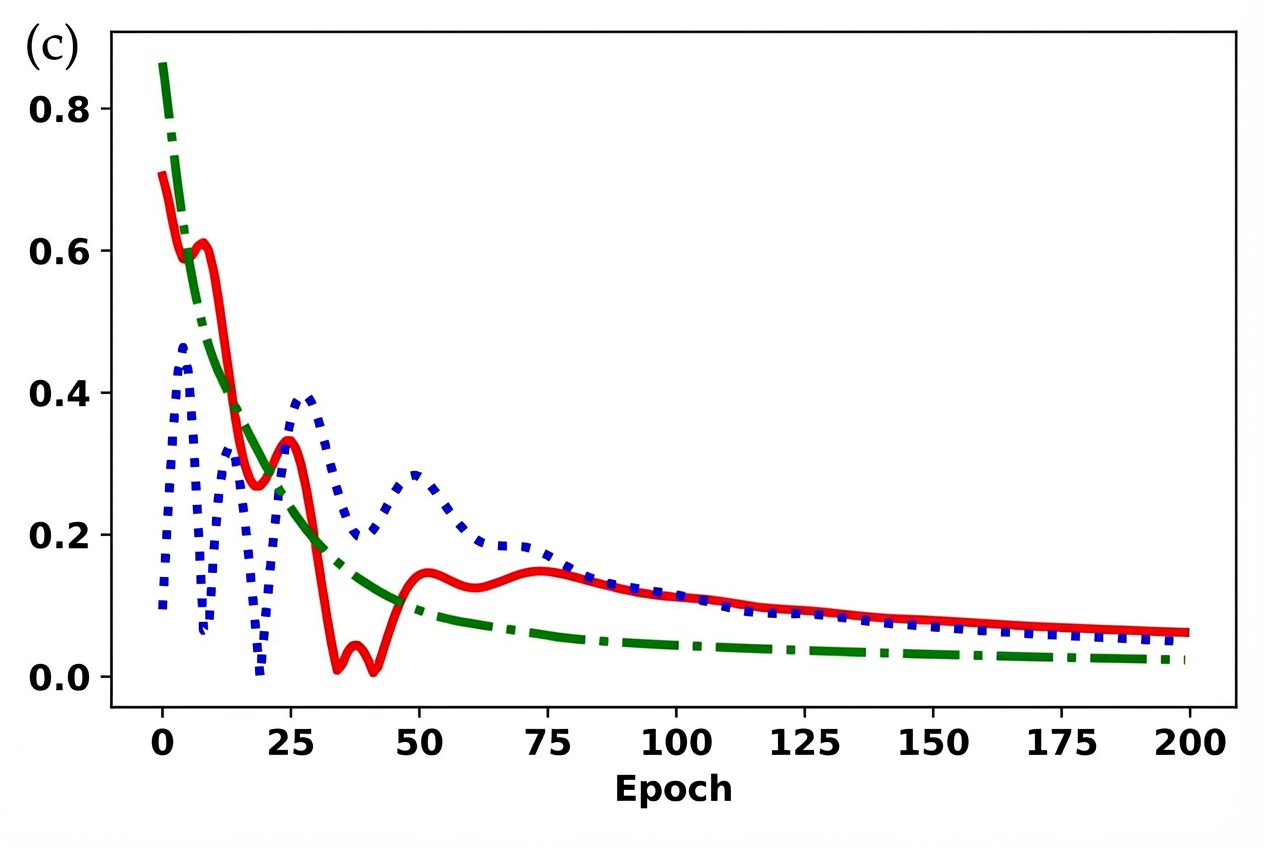}}
 \end{minipage}
	\caption{Evolutions of relative errors for low-frequency~(a), middle-~(b), and high-~(c) dominated functions over iterations.}\label{Fig:fre}
\vskip 0.1in
\end{figure*}

\subsubsection{Validation of Residual Behavior}
We contrast the analytical predictions derived from the QNTK regime with the actual training dynamics observed during the optimization of the three target functions defined in Eq.~\ref{eq:curves} to empirically validate our framework. 

As illustrated in Fig.~\ref{Fig:fit}, the analytical solutions exhibit a high degree of fidelity to the empirical residual dynamics. Remarkably, the residual dynamics verifies that the residuals of the frequency components possessing significant spectral amplitudes of the objective function decay exponentially under the setting of a small learning rate $\eta $. Moreover, the analytic solutions do not fit well with those of actual dynamics for the remaining, especially at the early stage of training. Nevertheless, the analytic approach for residual dynamics still gives a good description of the whole training process, since residuals for those remaining have a magnitude of several orders smaller than the residual of the frequency components with high spectral amplitude. 

Consequently, the proposed residual dynamics framework provides a robust approximation of the global training process. It  corroborates the spectral amplitude principle, showing that QNNs optimization do not inherently filter frequencies based on their indices, but rather prioritizes the rapid convergence of components with high spectral amplitudes.

\section{Experiments}\label{sec:exp}
To validate the proposed Spectral Amplitude Principle without the interference of hardware-specific noise and connectivity constraints which could potentially mask the intrinsic training dynamics we conduct our experiments in an ideal, noise-free environment. 

All the simulations are performed on a classical machine equipped with an Intel Core i7-14700HX CPU and an NVIDIA GeForce RTX 4060 Laptop GPU. We implement a Python quantum simulator simulate QNNs based on PennyLane~\cite{bergholm2018pennylane}.


\subsection{Preliminary Experiments}
We first examine the optimization trajectories of three univariate target functions to fit, as listed in Eq.~\ref{eq:curves} for a preliminary study. The quantum circuit comprises 4 qubits and a depth of 20 layers, with a total of 160 trainable parameters. 

\subsubsection{Example of Spectral Amplitude Priority}
Fig.~\ref{Fig:fre} illustrates the quantitative evidence of the frequency convergence behavior in QNNs, tracking the evolutions of relative errors for all frequency spectrum.

The results establish a decisive pattern: the rate of error convergence is primarily dictated by the magnitude of the spectral coefficients. As evidenced by the decay trajectories in Fig.~\ref{Fig:fre}, spectral components possessing significant amplitudes consistently exhibit the fastest convergence. This behavior is invariant to the frequency regime; in the case of the high-frequency dominated function (Fig.~\ref{Fig:fre} (c)), it prioritizes the high-frequency mode precisely as it carries the largest spectral weight. 
\begin{table}[t!]
    \centering
    \caption{Performance comparison.}
    \label{tab:model_performance}
    \resizebox{\linewidth}{!}{
        \begin{tabular}{lccc}
            \toprule
            \textbf{Config (Hidden Layers)} & \textbf{Params} & \textbf{Train Acc.} & \textbf{Test Acc.} \\
            \midrule
            $[200, 200, 200, 100]$ & 100,300 & 0.99 & 0.75 \\
            $[50, 50]$             & 2,600   & 0.63 & 0.62 \\
            $[100, 100]$           & 10,200  & 0.61 & 0.60 \\
            $[300, 300, 300]$      & 180,600 & 0.64 & 0.56 \\
            $[500, 500, 500, 500]$ & 751,000 & 0.87 & 0.70 \\
            \bottomrule
        \end{tabular}
    }
\end{table}
\begin{figure}[tb!]
\centering
 \includegraphics[width=0.42\textwidth]{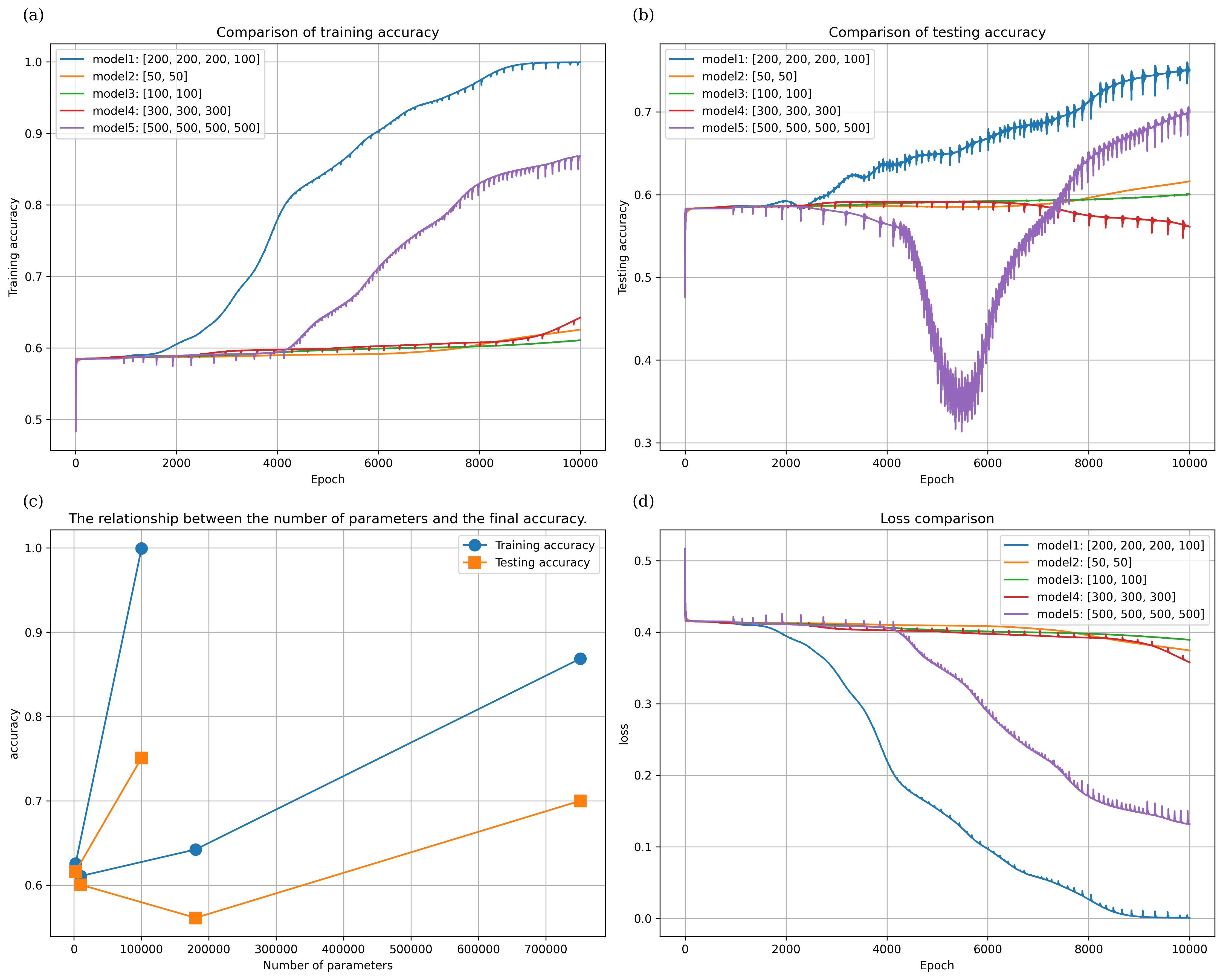}
 \caption{\textbf{Performance analysis of various DNN architectures on the high-frequency task.} (a-b) Evolution of training and test accuracy over 10K epochs for different configurations. (c) The relationship between model parameters and accuracy. (d) Comparison of training loss trajectories across the evaluated models.}\label{Fig:comparison}
\vskip 0.1in
\end{figure}
\begin{figure}[tb!]
	\centering
  \begin{minipage}{0.475\linewidth}
 	\centerline{\includegraphics[width=\textwidth]{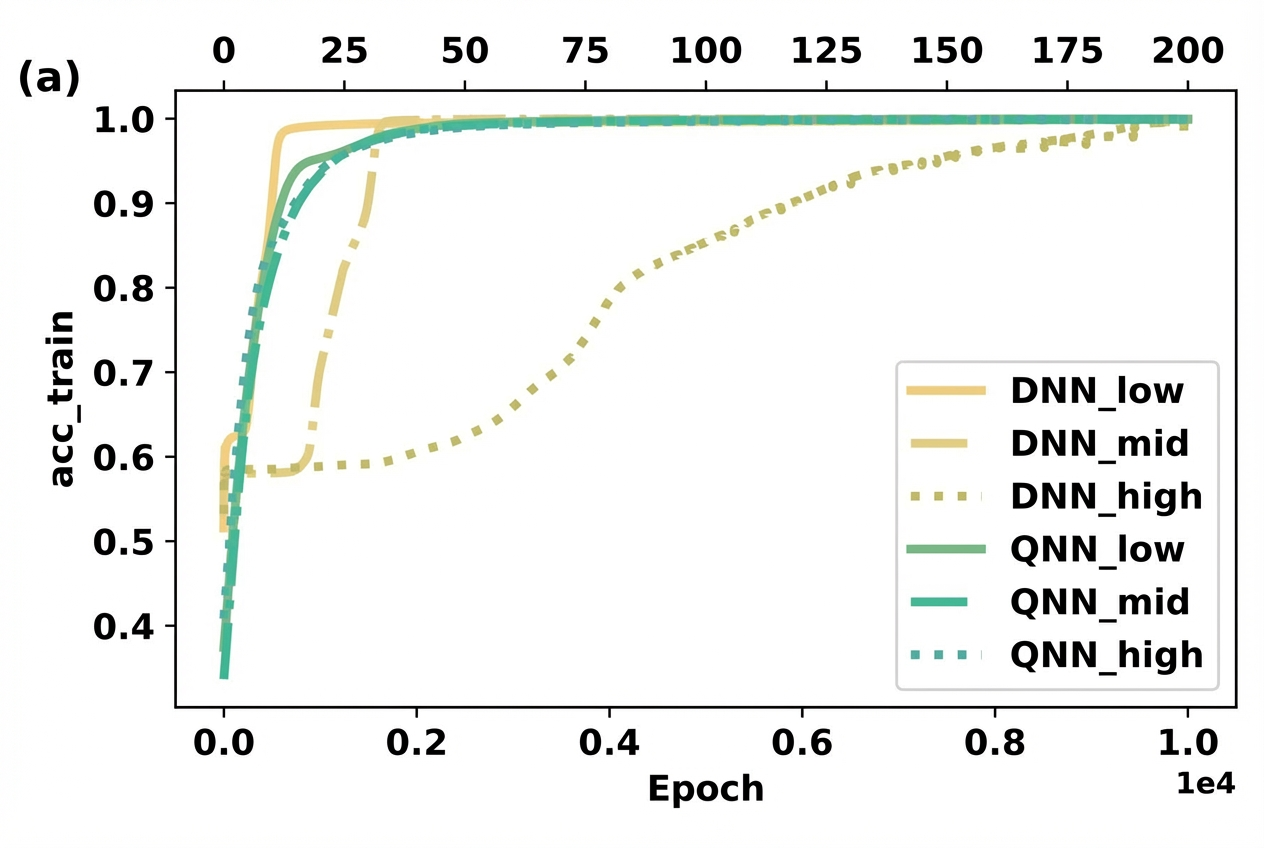}}
  \end{minipage}
  \begin{minipage}{0.475\linewidth}
     \centerline{\includegraphics[width=\textwidth]{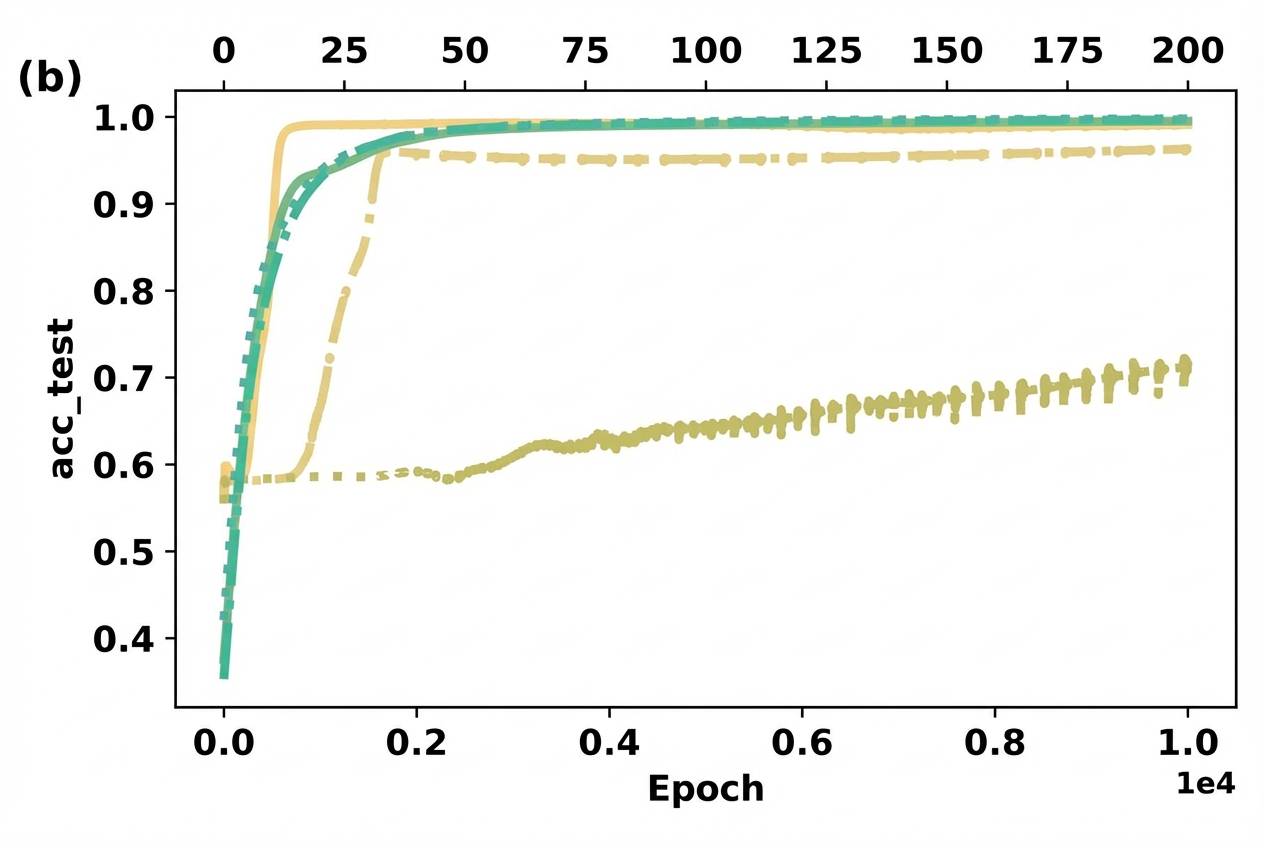}}
  \end{minipage}
	\caption{ \textbf{Comparison of accuracy of fitting~(a) and predicting~(b)  between DNNs and QNNs for low-frequency, middle-frequency and high-frequency dominated functions, respectively. } The horizontal axis, labeled ``Epoch", denotes the number of training iterations, while the vertical axis illustrates the training and testing accuracy of both DNNs and QNNs in the context of a one-variable function fitting task. In the x-coordinate, the bottom indicates the result of 10,000 times of DNN training and the top indicates 200 times of quantum circuit training.}
\label{Fig:acc}
\vskip 0.1in
\end{figure}
\begin{figure*}[tb!]
 \begin{minipage}{0.32\linewidth}
 	\centerline{\includegraphics[width=\textwidth]{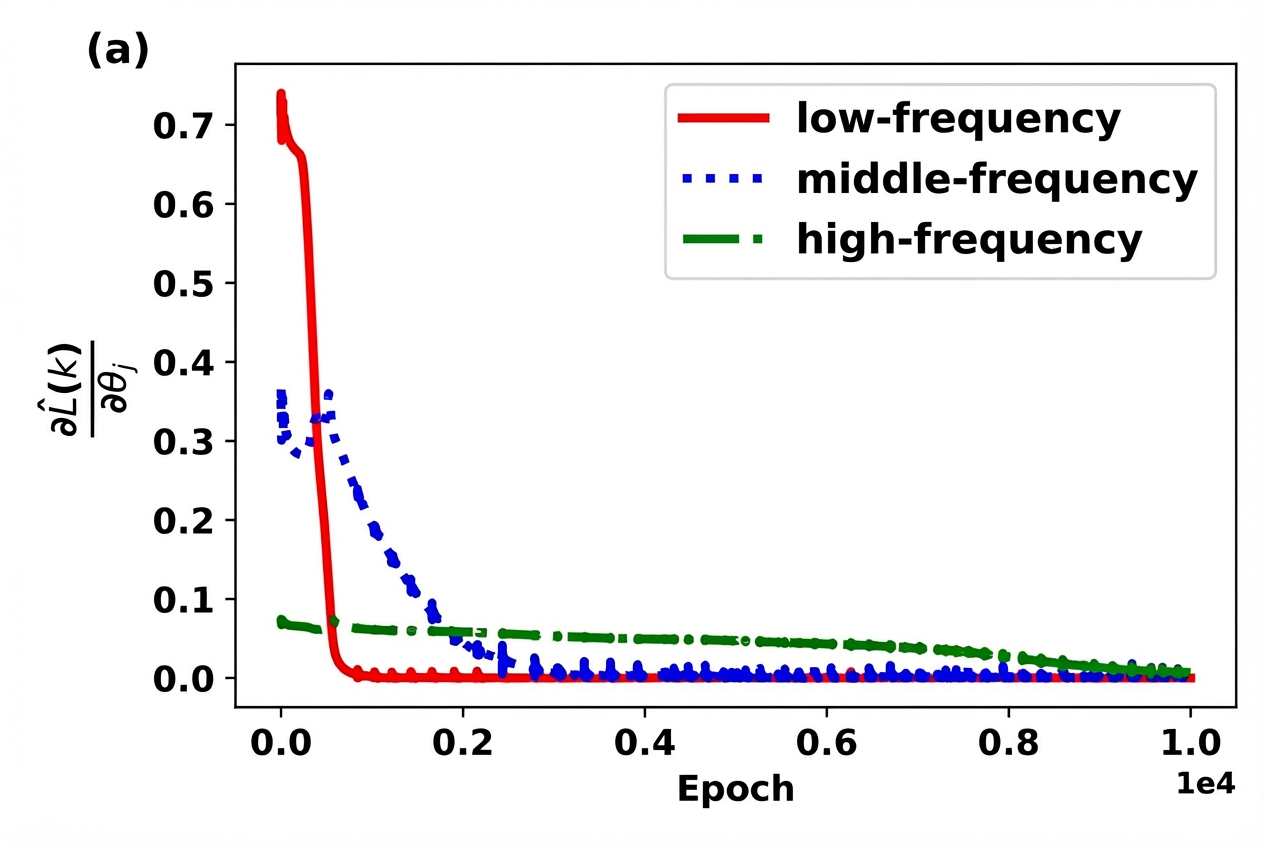}}
 \end{minipage}
 \begin{minipage}{0.32\linewidth}
     \centerline{\includegraphics[width=\textwidth]{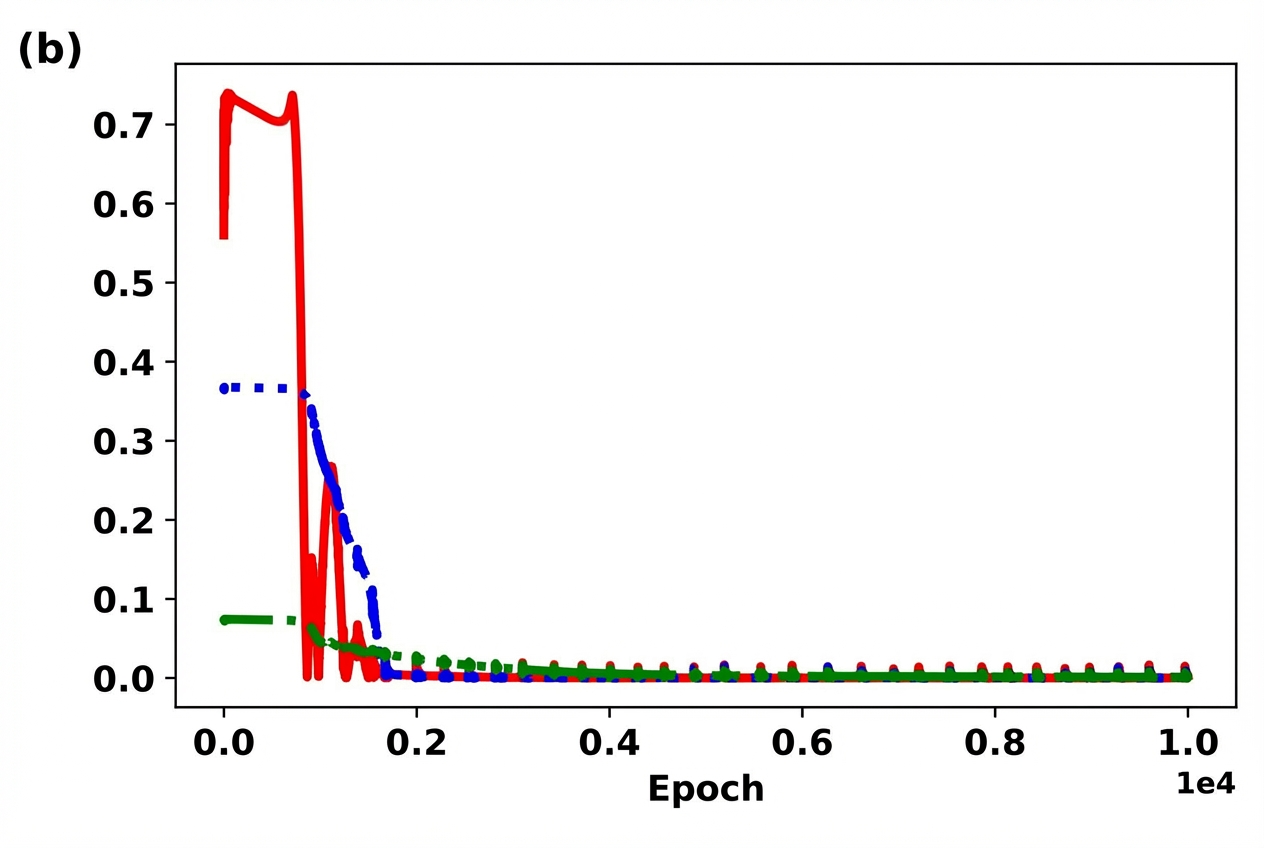}}
 \end{minipage}
  \begin{minipage}{0.32\linewidth}
     \centerline{\includegraphics[width=\textwidth]{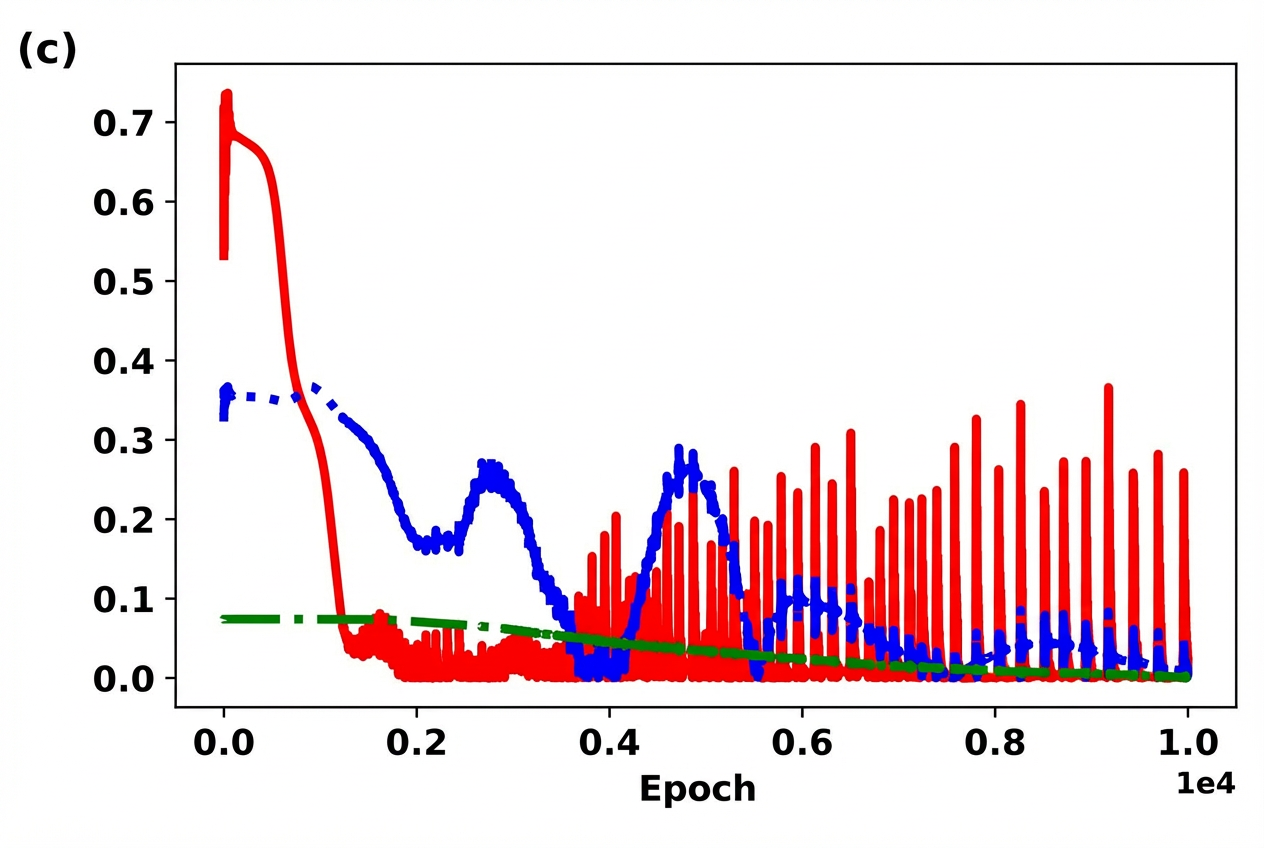}}
 \end{minipage}
	\caption{Evolutions of gradients for loss of DNNs in fitting low-frequency~(a), middle-~(b) and high- functions~(c).}\label{Fig:DNN}
\vskip 0.1in
\end{figure*}
\begin{figure}[tb!]
\centering
 \includegraphics[width=0.4\textwidth]{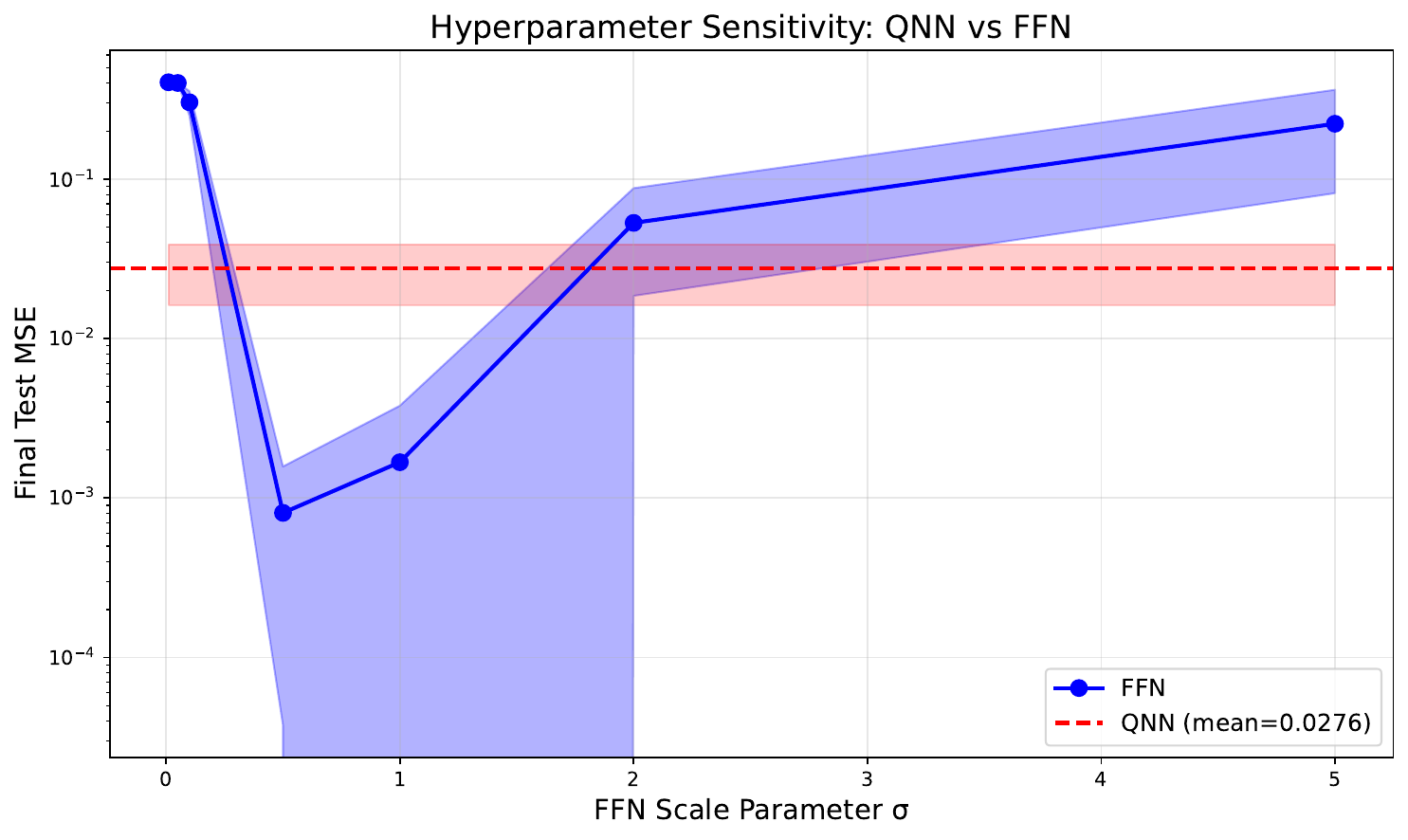}
 \caption{\textbf{Hyperparameter sensitivity comparison between QNN and FFN.} The plot illustrates the final test Mean Squared Error (MSE) as a function of the scale parameter for FFNs. The dashed red line represents the stable mean performance of the QNN, highlighting its robustness compared to the high sensitivity of classical FFNs to initialization scales.}\label{Fig:sensitivity}
\end{figure}

\subsubsection{Comparison with Classical DNNs}
We benchmark the spectral learning capabilities of QNNs against DNNs to validate the distinct inductive biases of the two architectures. To ensure a rigorous comparison, rather than selecting an arbitrary DNN, we perform an architectural sweep to identify the most competitive classical baseline. As summarized in Table~\ref{tab:model_performance}, Through an empirical comparison of models with parameter counts spanning several orders of magnitude ($2.6\times 10^3$ to $7.5\times10^5$), we find that the configuration [200, 200, 200, 100] yields the highest training (0.99) and testing (0.75) accuracy among the classical candidates. Notably, further increasing the model capacity (e.g., the [500, 500, 500, 500] configuration) does not result in superior performance, but rather leads to a noticeable degradation in test accuracy (0.70). This suggests that the difficulty in learning high-frequency functions cannot be simply mitigated by over-parameterization, highlighting a fundamental limitation of the inductive bias in classical architectures.  

The training and test dynamics for these varying DNN scales are further elucidated in Fig.~\ref{Fig:comparison}. The accuracy is measured by $1-L(\boldsymbol{\theta})$. Increasing the parameter budget fails to yield a proportional improvement in test accuracy for high-frequency function. Crucially, our comparative analysis in Table~\ref{tab:model_performance} and Fig.~\ref{Fig:comparison} focuses specifically on the high-frequency dominated function ($f_H(x)$). This choice is motivated by the fact that low- and middle-frequency components fall within the "comfort zone" of the classical spectral bias, where DNNs already exhibit efficient convergence. 

Above all, we choose the classical baseline which consists of a fully connected network with [200, 200, 200, 100] hidden layers (approx. $10^{4}$  parameters). As evidenced by the training and testing trajectories in Fig.~\ref{Fig:acc}, while the DNNs achieve high training accuracy across all target functions ($f_L(x)$, $f_M(x)$, $f_H(x)$) -- likely leveraging its massive parameter space for memorization, it fails to generalize on the high-frequency dominated task ($f_H(x)$). In contrast, despite possessing two orders of magnitude fewer parameters, the QNNs show robust generalization capabilities across the entire frequency spectrum. This suggests that the difficulty DNNs face with high-frequency functions is not merely a capacity issue but a fundamental limitation of their inductive bias, which QNNs  circumvent.

To rigorously characterize the mechanism behind this discrepancy, we analyze the frequency-domain evolution of the loss gradients for the DNNs, as illustrated in Fig.~\ref{Fig:DNN}. The gradient dynamics confirm that the classical model adheres strictly to the spectral bias, where optimization is preferentially driven by low-frequency components. Even when the target function is dominated by high frequencies ($f_H(x)$, Fig~\ref{Fig:DNN} (c)), the DNNs suppress high-frequency gradients, delaying their convergence. This stands in sharp contrast to the QNN dynamics (previously shown in Fig.~\ref{Fig: L_K}), where convergence is governed by spectral amplitude priority.
\subsubsection{Comparison with Fourier Feature Nets}
We compare QNNs against Fourier Feature Networks (FFNs), a prevalent classical architecture designed to mitigate spectral bias by mapping input data to a higher-dimensional feature space using random Fourier features. While FFNs can theoretically capture high-frequency components, their efficacy is heavily contingent upon the manual calibration of the scale parameter 
$\sigma $.

As illustrated in Fig.~\ref{Fig:sensitivity}, we evaluate the hyperparameter sensitivity of both architectures. the shaded regions denote the standard deviation (±std) computed over multiple independent trials with different random seeds, characterizing the training stability of both models. The FFN exhibits a pronounced ``U-shaped" sensitivity curve: an inadequately small $\sigma $ (e.g., 0.01) prevents the model from capturing high-frequency details, while an excessively large $\sigma $ (e.g., 5.0) leads to spectral aliasing and poor generalization. In stark contrast, the QNN shows remarkable robustness, maintaining a consistently low loss without requiring the fine-tuning of frequency-scaling hyperparameters. This suggests that the Spectral Amplitude Principle allows QNNs to adaptively prioritize relevant spectral components.

\subsection{Experiments on Classification}
We then consider the classification of Iris dataset $\left \{\left ( x_{i} ,y_{i} \right ) \right \} _{i = 0}^{N-1} $. where data $x_i$ has been preprocessed~(filtering and normalization for instance) and $y_{i}\in \{ -1,1 \}$ is the label for two classes. Regarding the theoretical foundation of this advantage, it has been mathematically established that classically simulating the output probability distributions of Instantaneous Quantum Polynomial circuits—typically characterized by a Hadamard-Diagonal-Hadamard structure—requires exponential overhead. Conversely, quantum processors can efficiently implement these circuits by simply executing the corresponding gate layers, thereby showing a significant and provable quantum speedup. The Iris task is employed here not as a means to prove the model's inherent superiority, but rather as a practical benchmark to evaluate the performance of a model already known for its advantages.

We adopt an ansatz with $2$ qubits and $6$ layers. The Mean Squared Error (MSE) is used as the loss function. 

\begin{figure}[tb!]
    \centering
  \begin{minipage}{0.46\linewidth}
 	\centerline{\includegraphics[width=\textwidth]{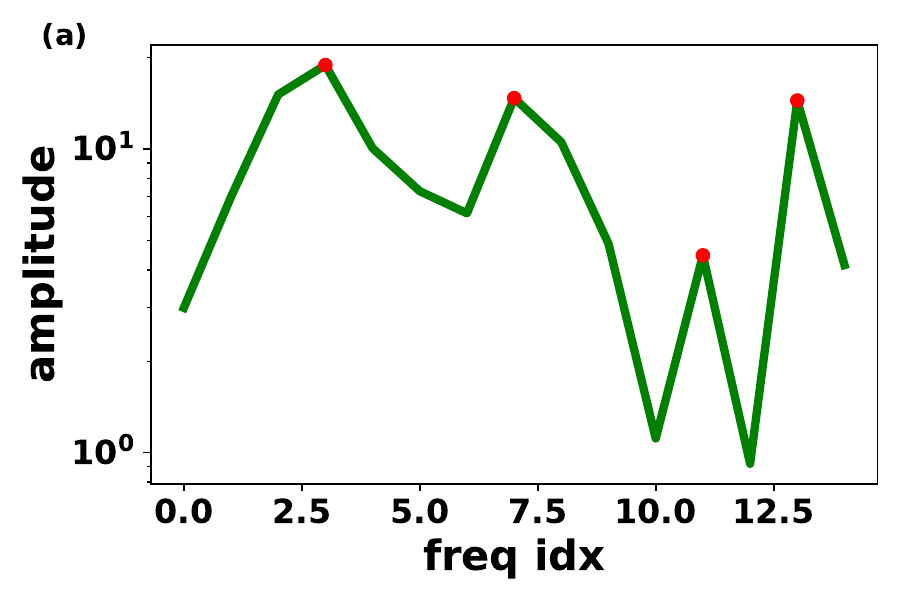}}
 \end{minipage}
 \begin{minipage}{0.42\linewidth}
     \centerline{\includegraphics[width=\textwidth]{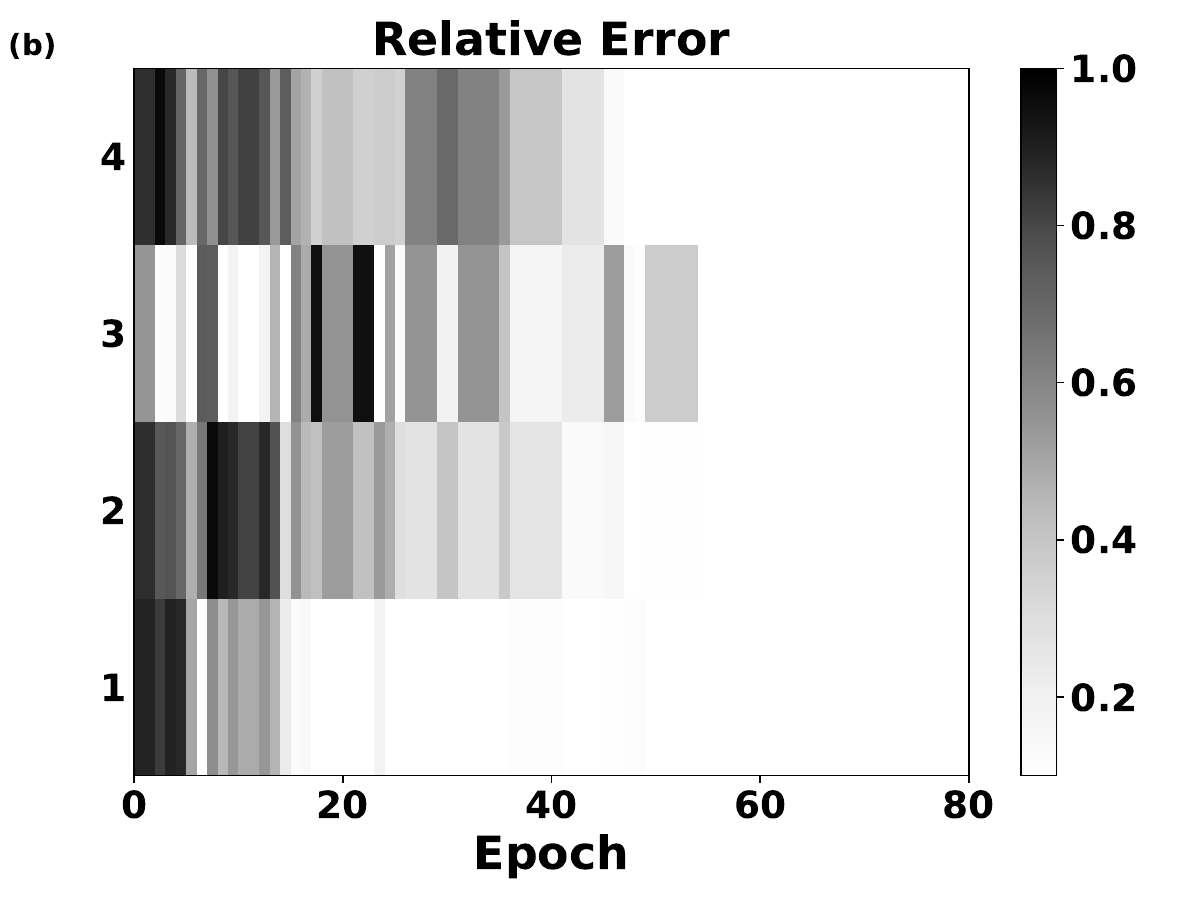}}
 \end{minipage}
    \caption{\textbf{Result of learning the Iris dataset with QNNs in the frequency domain.} (a) Magnitude-frequency map. (b) Evolution of relative errors at different frequencies during training. Vertical coordinates ‘1’, ‘2’, ‘3’, ‘4’ indicate the relative error of the selected frequency samples. The indices corresponding to the frequencies increase sequentially from the bottom to the top.}
    \label{Fig:iris}
\end{figure}

Fig.~\ref{Fig:iris} (a) presents the magnitude-frequency map of the target function, where the curve depicts the spectral distribution of the training data computed via the Discrete Fourier Transform. The spectrum reveals a significant non-uniformity in the distribution of spectral amplitudes across frequency components, characterized by a distinct peak structure. Red markers identify a set of representative frequency components, encompassing both the dominant frequencies with maximal amplitudes and secondary frequencies with lower magnitudes. This extraction of spectral features constitutes the foundation for the subsequent analysis; by defining the variations along the amplitude dimension, it serves as a critical reference for substantiating the correlation between fitting priority and spectral magnitude.

Fig.~\ref{Fig:iris} (b) shows the temporal evolution of relative errors for the selected frequency components during training. As a heatmap, the horizontal axis represents the training epochs, while the vertical axis corresponds to the specific frequency samples identified in Fig.~\ref{Fig:iris} (a). The color intensity encodes the magnitude of the relative error, ranging from black (indicating high error) to white (signifying low error).

\subsection{Experiments on Discrete Logarithm Problem}

We further perform experiments on the Discrete Logarithm Problem (DLP)~\cite{granger2018discrete}. The modular exponentiation inherent to the DLP exhibits extreme non-linearity, where the minute input perturbations induce drastic fluctuations within the solution space. From a spectral perspective, this volatility manifests as a distribution where the energy is predominantly dispersed across high-frequency components. Quantitatively, Fig. 11(a) reveals a spectral landscape where energy remains significant across high-frequency indices. This inherent lack of low-frequency structure presents a formidable challenge for DNNs, thereby establishing the DLP as an ideal testbed for validating the capacity of the Spectral Amplitude Priority mechanism. Remarkably, the quantum advantage of QNNs for DLP has been rigorously proved by using support vector machine armed with quantum kernel estimation~\cite{2020A}. In that work, it is showed that for datasets grounded in the DLP, any efficient classical algorithm is provably incapable of learning the classification rule, yielding performance indistinguishable from random guessing. Conversely, employing a support vector machine armed with quantum kernel estimation enables high-accuracy classification within polynomial time. While the original work uses a given quantum kernel, here we adopt a variational quantum circuit to parameterize the quantum kernel. This allows us to incorporate DLP tasks into the training framework of QNNs.

Specifically, we consider the DLP over the finite multiplicative group $Z_{p}^{\ast}$, defined by a large prime $p$ and a generator $\alpha$. The task is to recover the integer $x$ (where $0\le x\le p-2$) satisfying $\alpha ^{x} \equiv\beta$(mod $p$) denoted as $x = \log_{\alpha }\beta$. We conduct a labeled dataset $\left\{\left(x_{i},y_{i} \right)\right\}_{i = 0}^{N-1}$ based on discrete logarithmic properties.
\begin{figure}[tb!]
    \centering
  \begin{minipage}{0.46\linewidth}
 	\centerline{\includegraphics[width=\textwidth]{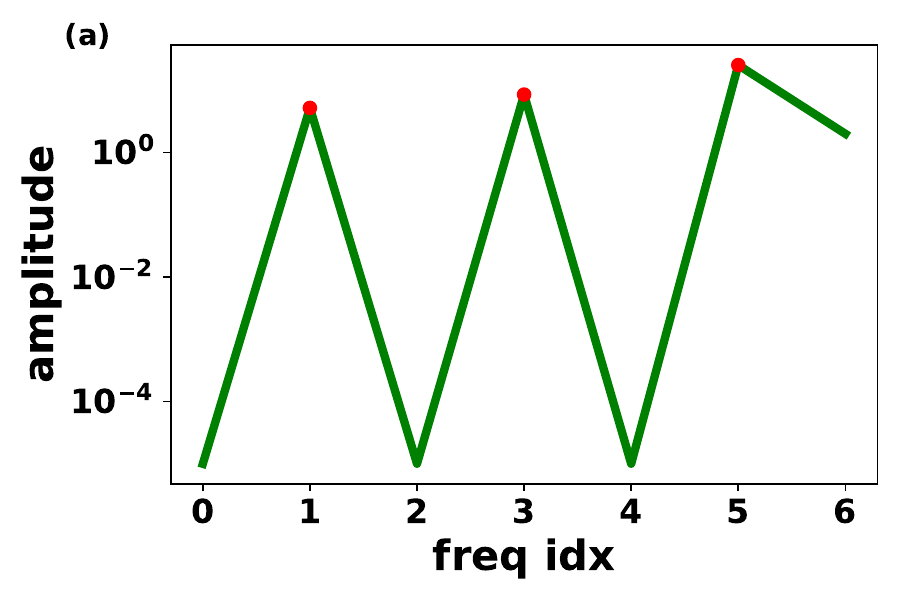}}
 \end{minipage}
 \begin{minipage}{0.42\linewidth}
 
     \centerline{\includegraphics[width=\textwidth]{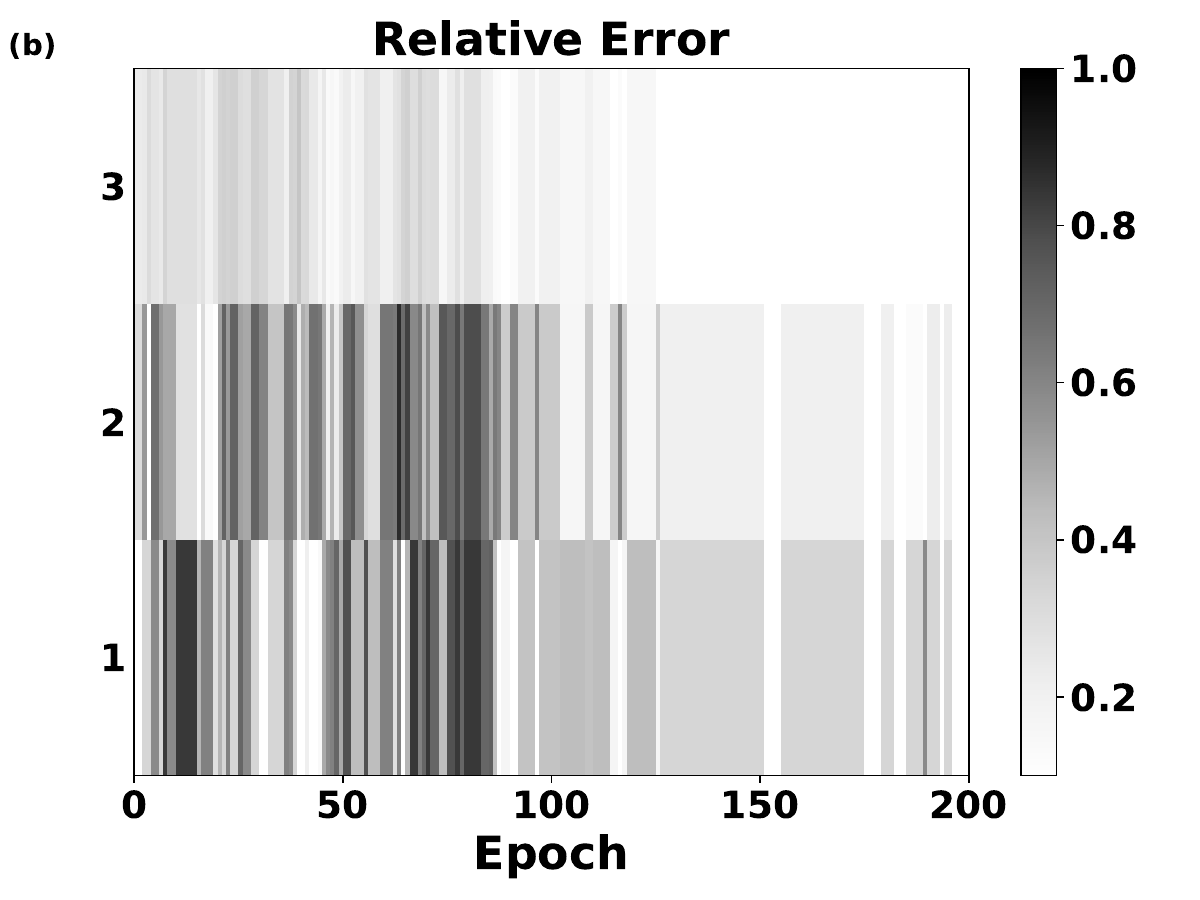}}
 \end{minipage}
    \caption{\textbf{Result of learning the discrete logarithm problem with QNNs in the frequency domain.} (a) Magnitude-frequency map. (b) Evolution of relative errors at different frequencies during training. The meanings of ‘1’, ‘2’ and ‘3’ in the vertical coordinates are the same as in Fig.~\ref{Fig:iris} (b).}
    \label{Fig:svm}
\vskip 0.1in
\end{figure}
Our experimental setup utilizes an 8 qubits and 24 layers ansatz with a data re-uploading architecture, applied to a dataset of 40 samples. We employ a variational quantum kernel framework where the input data $X$ is mapped to a high-dimensional Hilbert space via a parameterized ansatz $|\psi(x)\rangle=U(x,\theta)| 0\rangle^{\otimes n}$. The quantum kernel is defined as the overlap $K_{ij}=|\left\langle\psi(x_{i})|\psi(x_{j})\right\rangle|^{2}$. Optimization of the variational parameters $\theta$ is performed by maximizing kernel-target alignment~\cite{PhysRevA.106.042431} prior to SVM training. 

The training dynamics in frequency domain are governed in Fig.~\ref{Fig:svm}. The relevant definitions in both figures are analogous to those in Fig.~\ref{Fig:iris}. As shown in Fig.~\ref{Fig:svm} (b), the relative error decreases preferentially for frequencies with the highest spectral peaks (indicated by the red markers in Fig.~\ref{Fig:svm} (a)). Lower-amplitude components are subsequently learned in descending order of their spectral contributions. These results empirically confirm that Spectral Amplitude Priority governs QNN dynamics, enabling capturing of high-frequency components for quantum advantage.

\section{Conclusion}
\label{sec:level4} 
We have established the spectral amplitude priority as a fundamental mechanism characterizing the training dynamics of QNNs. We show that spectral fitting priority of QNNs is dictated by the magnitude of spectral coefficients rather than their frequency indices, which provides an explanation for the capability of QNNs to learn high-frequency data, overcoming the spectral bias that limits classical DNNs.

\nocite{langley00}

\bibliography{References}
\bibliographystyle{icml2026}

\newpage
\appendix
\onecolumn

\section{Related Work}\label{sec:level2}

The training dynamics of DNNs have evolved from empirical observations to rigorous theoretical formalizations. Early research has identified a pervasive ``Spectral Bias", where the classical neural networks prioritize low-frequency components during the early training phases~\cite{arpit2017closer,rahaman2019spectral}. To theoretically ground these observations, subsequent studies develop dynamic models such as the Linear Frequency-Principle~\cite{zhang2019explicitizing,luo2022exact}. This theoretical study culminates in the application of the Neural Tangent Kernel (NTK) framework, where convergence speeds are mathematically linked to the eigendecomposition of the kernel~\cite{basri2020frequency,ijcai2021p304} and further refined by analyzing the initialization distributions via partial differential equations~\cite{molina2024understanding}. Consequently, the spectral dynamics of classical models are now well-charted.

In the quantum domain, Fourier analysis has also emerged as a fundamental tool, primarily for characterizing the expressivity of PQC (namely QNN)~\cite{PhysRevA.103.032430}. Recent methodologies have focused on manipulating the frequency spectrum to enhance model capacity. Notable approaches include introducing trainable-frequency models to capture non-regularly spaced frequencies~\cite{jaderberg2024let}, defining metrics to evaluate ansatz capabilities in approximating Fourier series~\cite{heimann2025learning}, and developing selection algorithms to mitigate the curse of dimensionality in multi-dimensional inputs~\cite{poppel2025mitigating}.

However, these works primarily address the question of which frequencies are representable by the QNN, leaving the question of in what order frequencies are learned largely unexplored. Our work fills this gap by characterizing the spectral fitting (or convergence) dynamics during training. 

\begin{table}[tb!]
\centering
\caption{List of symbols and notations.}
\label{tab:symbols}
\footnotesize
\begin{tabular}{ll}
\toprule
\textbf{Symbol} & \textbf{Description} \\
\midrule
$x$ & Input data \\
$\boldsymbol{\theta}$ & Trainable parameters of QNN \\
$S(x)$ & Data encoding block \\
$W(\boldsymbol{\theta})$ & Trainable circuit block \\
$f(x, \boldsymbol{\theta})$ & QNN output function \\
$\hat{f}(k, \boldsymbol{\theta})$ & Fourier coefficient at frequency $k$ \\
$\Omega$ & Set of accessible frequencies \\
$\hat{\varepsilon}(k)$ & Residual in frequency domain \\
$A(k)$ & Amplitude of the residual \\
$\eta$ & Learning rate \\
$K(x, x')$ & QNTK \\
$\overline{K}(k, k')$ & Frequency-domain QNTK \\
\bottomrule
\end{tabular}
\end{table}

\end{document}